\documentclass[10pt, twocolumn]{extarticle}
\usepackage{multicol}
\usepackage[]{graphicx}
\usepackage[]{xcolor}
\usepackage{alltt}
\usepackage[T1]{fontenc}
\usepackage[utf8]{inputenc}
\usepackage{multirow}
\usepackage{gensymb}
\usepackage{subfigure}
\usepackage{amsmath}
\usepackage{amssymb}
\usepackage{cite}
\usepackage{array}
\usepackage[]{authblk}
\usepackage{threeparttable}
\usepackage{textcomp}
\usepackage{babel}
\usepackage[switch]{lineno}
\usepackage{cases} 
\usepackage{empheq}
\usepackage{appendix}
\usepackage{bm}
\usepackage{tabularx}
\usepackage{array}
\usepackage{adjustbox}
\usepackage{makecell}  
\usepackage{soul}
\usepackage{caption}
\makeatletter
\newcommand{\manualsuppref}[3]{%
  \@namedef{r@#1}{{#2}{}{}{}{}}%
  \@namedef{r@#1@cref}{{#3}{}{}{}{}}%
}

\manualsuppref{SSec_Dominant_defects}{S1}{[section][1][]S1}
\manualsuppref{Gfanti}{S1}{[equation][1][]S1}
\manualsuppref{Gftp}{S4}{[equation][4][]S4}
\manualsuppref{Canti1}{S5}{[equation][5][]S5}
\manualsuppref{Canti2}{S6}{[equation][6][]S6}
\manualsuppref{Ctp1}{S7}{[equation][7][]S7}
\manualsuppref{Ctp2}{S8}{[equation][8][]S8}
\manualsuppref{STab_point_combination}{S1}{[table][1][]S1}
\manualsuppref{STab_point_combination_conc}{S2}{[table][2][]S2}
\manualsuppref{sec:derivation}{S1.2}{[subsection][2][1]S1.2}
\manualsuppref{eq:fraction}{S19}{[equation][19][]S19}

\manualsuppref{SSec_decompos_delta_G_f}{S2}{[section][2][]S2}
\manualsuppref{stab_deltaG_thermal_3000K}{S3}{[table][3][]S3}

\manualsuppref{SSec_thermal_properties}{S3}{[section][3][]S3}
\manualsuppref{STab_0K_properties}{S4}{[table][4][]S4}
\manualsuppref{sfig_TherPro}{S1}{[figure][1][]S1}
\manualsuppref{sfig_DOSs}{S2}{[figure][2][]S2}
\manualsuppref{sfig_DOSs-finite_T}{S3}{[figure][3][]S3}

\manualsuppref{SSec_MTP_training}{S4}{[section][4][]S4}
\manualsuppref{STab_MTP-training}{S5}{[table][5][]S5}

\manualsuppref{SSec_Comp_Para}{S5}{[section][5][]S5}
\manualsuppref{SSSec_FES_para}{S5.1}{[subsection][1][5]S5.1}
\manualsuppref{Seq_Fah}{S21}{[equation][21][]S21}
\manualsuppref{Seq_integration}{S22}{[equation][22][]S22}
\manualsuppref{Seq_upsampling}{S23}{[equation][23][]S23}
\manualsuppref{Seq_F_up+el}{S24}{[equation][24][]S24}
\manualsuppref{Seq_Fel}{S25}{[equation][25][]S25}

\manualsuppref{tab:FES_common}{S6}{[table][6][]S6}
\manualsuppref{tab:FES_bulk_up}{S7}{[table][7][]S7}
\manualsuppref{tab:FES_path}{S8}{[table][8][]S8}
\manualsuppref{STab_para_DFT_0K_qh}{S9}{[table][9][]S9}
\manualsuppref{STab_up_para_bulk}{S10}{[table][10][]S10}
\manualsuppref{STab_up_para_defects}{S11}{[table][11][]S11}
\manualsuppref{STab_el_para_defects}{S12}{[table][12][]S12}
\manualsuppref{STab_para_Pure}{S13}{[table][13][]S13}

\manualsuppref{SSec_Conv_tests}{S5.4}{[subsection][4][5]S5.4}
\manualsuppref{sfig_convergence_test}{S4}{[figure][4][]S4}
\manualsuppref{STab_para_MTP2}{S14}{[table][14][]S14}

\manualsuppref{SSec_MTP2}{S6}{[section][6][]S6}
\manualsuppref{sfig_DFT_VS_MTP}{S5}{[figure][5][]S5}

\manualsuppref{SSec_mu_A2}{S7}{[section][7][]S7}
\manualsuppref{STab_para_DFT_0K_A2}{S15}{[table][15][]S15}

\makeatother

\newcommand{\Ganti}{G_{\mathrm{f, anti}}}
\newcommand{\Canti}{C_{\mathrm{anti}}}

\usepackage[colorlinks,
			linkcolor=blue,
			anchorcolor=green,
			citecolor=blue,
			urlcolor=blue
			]{hyperref}
\usepackage[nameinlink,noabbrev]{cleveref} % <-- new
\crefname{equation}{}{} % capitalize "E", no period

\usepackage{fancyhdr}
\usepackage[a4paper,top=50pt,bottom=50pt,left=50pt,right=50pt]{geometry}

\usepackage{textcomp}

\def\keywords{\vspace{.3em}
{\textit{Keywords}:\,\relax%
}}

\usepackage{float}
\floatstyle{plaintop}
\restylefloat{table}

\author[1]{Xiang Xu}
\author[2,3]{Fritz K$\rm{\ddot{o}}$rmann}
\author[4]{Sergiy Divinski}
\author[1]{Blazej Grabowski}
\author[1,*]{Xi Zhang}

\affil[1]{Institute for Materials Science, University of Stuttgart, Pfaffenwaldring 55, 70569 Stuttgart, Germany}
\affil[2]{Interdisciplinary Centre for Advanced Materials Simulation, Ruhr-Universit{\"a}t Bochum, 44801 Bochum, Germany}
\affil[3]{Department for Computational Materials Design, Max-Planck-Institut for Sustainable Materials,  Max-Planck-Str.1, 40237 D$\ddot{u}$sseldorf, Germany}
\affil[4]{Institute of Materials Physics, University of Münster, Wilhelm-Klemm-str. 10, 48149 Münster, Germany}
\affil[*]{\rm E-mail: xi.zhang@imw.uni-stuttgart.de}

\date{}
\title{Temperature dependence of the Gibbs energies of formation of point defects in B2 MoTa from \textit{ab initio} calculations}

\begin{document}

\twocolumn[
\begin{@twocolumnfalse}
\maketitle
\vspace{-1cm}

\section*{Abstract}
Using B2 MoTa---the strongest B2 former among group V/VI refractory binaries---as a model system, we compute temperature-dependent Gibbs energies of formation of vacancies and antisites from \textit{ab initio} calculations up to 3000~K at the stoichiometric composition. We explicitly account for thermal electronic excitations, vibrational anharmonicity, and electron--vibration coupling. The key finding is that the temperature dependence of the Gibbs energies of vacancy formation exhibits a pronounced sublattice asymmetry. Specifically, the Gibbs energy of formation of a Mo-site vacancy decreases by 1.1 eV from 0 to 3000 K, whereas the decrease for a Ta-site vacancy amounts to 2.1 eV, almost a factor of two larger. 
Two contributions of distinct origin govern the temperature dependence of this asymmetry: a quasiharmonic contribution associated with the chemical-potential imbalance set by the two antisite structures and an anharmonic contribution associated with the local vibrational response of the vacancy structures.
The asymmetry in anharmonic vibrations is traced back to an enlarged local vibrational distribution of the first-nearest neighbors around the Ta-site vacancy. 
In contrast to the vacancies, the Gibbs energies of antisite formation vary only weakly with temperature.

\keywords Refractory alloys; B2 compounds; Vacancy thermodynamics; Sublattice asymmetry; Vibrational effects.

\vspace{0.5cm}
\end{@twocolumnfalse}
]

\begin{sloppypar}

\section{Introduction}

Point defects govern atomic diffusion and thereby influence diffusion-mediated microstructural evolution, phase stability, and phase transformations~\cite{shewmon2016diffusion}.
Point-defect formation energies are essential for describing diffusion in systems ranging from elemental metals~\cite{angsten2014elemental} to intermetallic compounds~\cite{mishin1997atomistic,xu2011effect} and multicomponent alloys~\cite{dash2022recent}.
In chemically disordered alloys, these energies depend sensitively on the local chemical environment~\cite{zhang2022ab,linton2025mechanistic,luo2025determinants}, producing wide energy distributions that challenge the quantitative prediction, especially at elevated temperatures~\cite{dash2022recent}.
Ordered intermetallic compounds, in contrast, provide a well-defined reference system in which sublattice-dependent defect energetics can be quantified and related to the broader problem of environment-dependent defect behavior in chemically complex solid solutions.
The B2 structure is a particularly useful prototype in this context, as its two-sublattice order allows clear assignment of defects to specific sublattices. 
However, datasets that resolve defect-formation Gibbs energies at elevated temperatures, including vibrational and electronic thermal contributions, remain limited for ordered intermetallic compounds.

This limitation is especially critical for body-centered cubic (BCC) refractory alloys used in high-temperature applications such as aerospace, power generation, and nuclear systems~\cite{pickering2021high,zhuo2024review,zhang2025refractory}.
In these systems, B2-type chemical ordering or A2/B2 phase coexistence is frequently reported and can significantly modify the high-temperature performance~\cite{zhuo2024review}.
Among group~V/VI refractory binaries, the tendency for B2 ordering is strongest in the Mo--Ta system~\cite{wang2024stability, blum2004mixed, blum2005prediction}.
Mo--Ta nearest-neighbor correlations also dominate characteristic short-range order motifs in multicomponent MoTaNbW alloys~\cite{woodgate2023short, zhang2022phase}.
In this sense, B2 MoTa represents a well-defined limiting case of chemical ordering, offering a controlled reference to disentangle sublattice-dependent defect energetics that are likewise relevant to chemically more complex refractory alloys.
Experimentally, however, equilibrated B2 MoTa with long-range order is difficult to obtain, likely due to kinetic constraints associated with intrinsically slow diffusion near the transition temperature.

Given the limited experimental access to equilibrated B2 MoTa, theoretical thermodynamic descriptions play a central role.
\textit{Ab initio} calculations have established reliable 0~K energetics for defect-free B2 and A2 MoTa~\cite{wang2018computation, blum2004mixed, turchi2005application}, yet quantitative finite-temperature Gibbs energies for point defects, resolved by defect type and sublattice, remain limited.
For alloy design, CALPHAD (CALculation of PHAse Diagrams) approaches provide a complementary route to finite-temperature phase stability~\cite{turchi2005application, zhang2022phase}, and recent advances in defect-energy formalisms have established clear links between point-defect energetics and CALPHAD Gibbs energy parameters~\cite{orvati2024defect}.
These developments highlight the need for accurate \textit{ab initio} formation Gibbs energies of point defects in ordered compounds for which experimental thermodynamic data are scarce, particularly in refractory systems at high temperatures where electronic and vibrational thermal excitations can be substantial~\cite{forslund2023thermodynamic,srinivasan2023anharmonicity}.

A systematic finite-temperature description of point defects in ordered intermetallics must explicitly account for their sublattice dependence and thermodynamic complexity.
Theoretical frameworks have been proposed for point-defect thermodynamics in B2 aluminides, including the determination of the antisite-dominated and triple-defect mechanisms~\cite{mishin1997atomistic,meyer1999atomic, leitner2015thermodynamics}, the consideration of concentrated point defects~\cite{xu2011effect}, and the influence on phase stability~\cite{goiri2016phase}.
In contrast, this topic has been only briefly addressed for refractory B2 compounds~\cite{medasani2016predicting}, and a systematic investigation is still lacking.
Moreover, the influence of lattice vibrations on the Gibbs energies of defect formation at elevated temperatures has been sparsely explored in ordered intermetallics via the quasiharmonic or frozen-phonon descriptions~\cite{lozovoi2003point,tingaud2014point}, without considering anharmonic vibrations.
This gap is notable in view of \textit{ab initio} studies on elemental metals, such as face-centered cubic (FCC) Ni~\cite{gong2018Nivac}, FCC Cu~\cite{glensk2014PRX}, and BCC W~\cite{zhang2025ab}, which demonstrate that anharmonic vibrational effects can significantly modify vacancy formation Gibbs energies.

Here, we investigate stoichiometric B2 MoTa up to high temperatures.
We determine the temperature dependence of the Gibbs energies of formation of vacancies and antisites using an \textit{ab initio} framework that combines thermodynamic integration and free-energy perturbation techniques~\cite{zhou2022thermodynamics, jung2023high}.
Thermal electronic excitations and explicit vibrational anharmonicity are included up to 3000~K, near the MoTa solidus (3119~K from experiment~\cite{rajkumar2020measurements}). 
Although long-range B2 order in MoTa disappears below this temperature, calculations up to 3000~K provide an upper bound for the assessment of thermal electronic and vibrational contributions to defect formation Gibbs energies, which are crucial for CALPHAD-type modeling of alloys exhibiting B2-type ordering or A2/B2 coexistence.
To enable efficient sampling of the relevant configurational and vibrational phase space, a machine-learning-based moment tensor potential (MTP) is employed~\cite{shapeev2016moment, novikov2020mlip}.
Based on the resulting temperature-dependent Gibbs energies of formation, equilibrium defect populations are evaluated within the dilute-solution approximation. 

\section{Methodology}
\subsection{Point-defect energetics in B2 MoTa}\label{Sec_Method_B2}

Within a typical density functional theory (DFT)-based supercell approach, the defect formation Gibbs energy $G_{\mathrm{f}, d}(T)$ of a defect $d$ can be written as~\cite{zhang2018calculating} (omitting the pressure dependence for brevity)
\begin{equation}\label{EQ_PDE}
G_{\mathrm{f}, d}(T) = G_{d}(T) - G_{\mathrm{perf}}(T) - \sum_i \Delta n_i \mu_i(T),
\end{equation} 
where $G_{d}(T)$ and $G_{\mathrm{perf}}(T)$ are respectively the Gibbs energy of a supercell with and without the considered defect (see Table~\ref{tab_defects}).
Further, $\Delta n_i$ is the change in the number of atoms of element $i$ ($i$ standing for Mo or Ta in the present work), with $\Delta n_i >0$ ($\Delta n_i<0$) representing the addition (removal) of atoms to (from) the supercell. Finally in Equation~\eqref{EQ_PDE}, $\mu_i$ is the chemical potential of element $i$, which is related to the Gibbs energy of a perfect B2 crystal as
\begin{equation}\label{EQ_constr_1}
    G_{\mathrm{perf}}(T) = \sum_i N_i \mu_i(T),
\end{equation}
where $N_i$ is the number of atoms of element $i$, summing up to the total number of atoms in the supercell, i.e., $N=\sum_i N_i$.

\begin{table*}[t]
    \centering
    \begin{tabular}{c|cccl|c}
    \hline
         Defect & $d$ &  $\Delta n_{\mathrm{Ta}}$ & $\Delta n_{\mathrm{Mo}}$ & Description & $E_{{\mathrm{f}}, d}^{\mathrm{0K}}$~(eV)\\
         \hline
         Mo antisite & $\mathrm{Mo_{Ta}}$ & $-1$ & $\phantom{-}1$ & Mo on Ta sublattice & $0.30$ \\
         Ta antisite & $\mathrm{Ta_{Mo}}$ & $\phantom{-}1$ & $-1$ & Ta on Mo sublattice & $0.30$ \\
         Mo vacancy & $\mathrm{vac_{Mo}}$ & $\phantom{-}0$ & $-1$ & vacancy on Mo sublattice & $3.06$ \\
         Ta vacancy & $\mathrm{vac_{Ta}}$ & $-1$ & $\phantom{-}0$ & vacancy on Ta sublattice & $4.49$ \\ \hline
    \end{tabular}
    \caption{Considered point defects $d$ in B2 MoTa and their formation energies at 0~K, $E_{{\mathrm{f}}, d}^{\mathrm{0K}}$. The change in the number of Ta (Mo) atoms in the supercell is denoted as $\Delta n_{\mathrm{Ta}}$ ($\Delta n_{\mathrm{Mo}}$). 
    }
    \label{tab_defects}
\end{table*}

Computing the chemical potential $\mu_i(T)$ requires special consideration in ordered intermetallics. 
It is necessary to determine the dominant point defect type that influences the chemical potential~\cite{mishin1997atomistic,meyer1999atomic}.
Comparison of the formation energies of the relevant point-defect combinations (see Section~\ref{SSec_Dominant_defects} in the Supplementary Materials for details) confirms that antisite pairs are the dominant point-defect combinations in B2 MoTa, in agreement with Ref.~\cite{medasani2016predicting}.
As a consequence, the concentrations of the two types of antisites, $C_{\mathrm{Mo}_{\mathrm{Ta}}}$ and $C_{\mathrm{Ta}_{\mathrm{Mo}}}$, can be assumed to be similar in stoichiometric B2 MoTa,
\begin{equation}\label{EQ_constr_2}
    C_{\mathrm{Mo}_{\mathrm{Ta}}}(T) \approx C_{\mathrm{Ta}_{\mathrm{Mo}}}(T).
\end{equation}
In the dilute limit, Equation~\eqref{EQ_constr_2} suggests that the formation energies of the two antisites can be treated as equal,
\begin{equation}
 G_{\mathrm{f,Mo_{Ta}}}(T) = G_{\mathrm{f,Ta_{Mo}}}(T) \equiv \Ganti(T), \label{EQ_constr_3}
\end{equation}
where $\Ganti(T)$ denotes the common (effective) Gibbs energy of antisite formation.
Then the chemical potential of B2 MoTa can be obtained from Equations~\eqref{EQ_PDE}, \eqref{EQ_constr_1} and \eqref{EQ_constr_3}:
\begin{align}
    \mu_{\mathrm{Mo}}(T)=\frac{G_{\mathrm{perf}}(T)}{N}+\frac{G_{\mathrm{Mo_{Ta}}}(T)-G_{\mathrm{Ta_{Mo}}}(T)}{4}, \label{EQ_mu_Mo}\\
    \mu_{\mathrm{Ta}}(T)=\frac{G_{\mathrm{perf}}(T)}{N}-\frac{G_{\mathrm{Mo_{Ta}}}(T)-G_{\mathrm{Ta_{Mo}}}(T)}{4}. \label{EQ_mu_Ta}
\end{align}
With the chemical potentials available, the Gibbs energy of defect formation can be obtained from Equation~\eqref{EQ_PDE}. From the temperature dependence of $G_{\mathrm{f}, d}$, the entropy of defect formation can be calculated as 
\begin{equation}\label{EQ_entropy}
    S_{\mathrm{f}, d}(T) = -\frac{\partial G_{\mathrm{f}, d}(T)}{\partial T}.
\end{equation}
In the dilute limit, the sublattice site-fraction of a point defect $d$ is given by
\begin{equation}
    X_{d} (T) = \exp\left(-\frac{G_{\mathrm{f},d}(T)}{k_{\mathrm{B}}T}\right),
    \label{EQ_Concent}
\end{equation}
with $k_{\mathrm{B}}$ the Boltzmann constant. 
The global concentration for a point defect $d$ in a B2 compound is then defined as 
\begin{equation}
    C_{d} (T) = \frac{1}{2} X_{d}(T).
    \label{EQ_Concent_Globle}
\end{equation}

\subsection{Free energy contributions}\label{Sec_Method_free_energy}

The Gibbs energy can be obtained from the Helmholtz free energy by a Legendre transformation.
Within the adiabatic approximation, the Helmholtz free energy can be decomposed as
\begin{equation}\label{EQ_Helm_F}
    F(V, T) = E^{\mathrm{0K}}(V) + F^{\mathrm{vib}}(V, T) + F^{\mathrm{el}}(V, T).
\end{equation}
Here, the superscripts denote the individual energetic contributions: $E^{\mathrm{0K}}$ denotes the static 0~K energy, while $F^{\mathrm{vib}}$ and $F^{\mathrm{el}}$ represent the vibrational and thermal electronic contributions, respectively, with the electron-vibration coupling included in $F^{\mathrm{el}}$. The vibrational contribution can be written as
\begin{equation}
    F^{\mathrm{vib}}(V, T) = F^{\mathrm{qh}}(V, T) + F^{\mathrm{ah}}(V, T),
\end{equation}
where $F^{\mathrm{qh}}$ is the quasiharmonic free energy and $F^{\mathrm{ah}}$ the explicit anharmonic free energy. This decomposition of the Helmholtz free energy leads to an analogous decomposition of the Gibbs energy of formation:
\begin{equation}
\begin{split}
    G_{\mathrm{f},d}(T) = & E^{\mathrm{0K}}_{\mathrm{f},d} + G^{\mathrm{qh}}_{\mathrm{f},d}(T) + G^{\mathrm{ah}}_{\mathrm{f},d}(T) + G^{\mathrm{el}}_{\mathrm{f},d}(T).
\end{split}
\end{equation}

All reported contributions are evaluated at the level of DFT accuracy. For $F^{\mathrm{qh}}$, the finite displacement method implemented in Phonopy~\cite{phonopy} is used, based on DFT forces. 
For $F^{\mathrm{ah}}$, the direct-upsampling method is used \cite{jung2023high}. In brief, the quasiharmonic reference is used for thermodynamic integration to a specifically optimized MTP (Equation~\eqref{Seq_integration}). DFT accuracy is recovered by free-energy perturbation theory, i.e., by upsampling to DFT energies (Equation~\eqref{Seq_upsampling}). In $F^{\mathrm{ah}}$, the electronic temperature is set to 0~K (using the Methfessel-Paxton smearing) to isolate the purely anharmonic contribution. The electronic free energy, $F^{\mathrm{el}}$, including the coupling to atomic vibrations, is obtained from Mermin's finite-temperature formalism \cite{mermin1965thermal}, applied to the upsampled snapshots (Equations~\eqref{Seq_F_up+el} and \eqref{Seq_Fel}).

In the present work, we do not compute full free-energy surfaces for all supercells. To reduce the computational cost, we instead follow a modified approach that still captures the dominant physics. For the bulk supercell, the vibrational free-energy surface is computed at the DFT level to determine the thermal expansion at zero pressure, $V_{\mathrm{eq, bulk}}(T)$. The Gibbs energy of defect formation, $G_{\mathrm{f},d}(T)$, is then approximated from Helmholtz free-energy differences between the defect and bulk supercells evaluated along $V_{\mathrm{eq, bulk}}(T)$. The electronic free-energy contribution is included along the same expansion path.

\subsection{Computational details}\label{Sec_Method_Comp_para}

A B2 supercell containing 128 lattice sites is used as the reference defect-free supercell.
Defect supercells are generated by introducing one of the four point defects defined in Table~\ref{tab_defects}.
Accordingly, antisite supercells contain 128 atoms, whereas vacancy supercells contain 127 atoms.
Calculations using a larger 250-atom supercell (Figure~\ref{sfig_convergence_test}(c) in the Supplementary Materials) confirm that the 0~K defect formation energies are converged with respect to the supercell size.

DFT calculations are performed using the Vienna \emph{ab initio} Simulation Package (VASP)~\cite{kresse1993ab,kresse1994ab} with projector augmented-wave (PAW) potentials~\cite{blochl1994projector} and the Perdew-Burke-Ernzerhof (PBE) generalized-gradient approximation (GGA)~\cite{perdew1996generalized} for the exchange-correlation functional.
The utilized PAW potentials treat the $4p^65s^14d^5$ (Mo$_{\mathrm{pv}}$) and $5p^66s^25d^3$ (Ta$_{\mathrm{pv}}$) electrons as valence for Mo and Ta, respectively. 
For DFT calculations at 0~K and for the upsampling configurations, the default plane-wave energy cutoff of 224.6~eV and a $k$-point mesh of $4\times4\times4$ are used.
For phonon calculations, a higher plane-wave energy cutoff of 500~eV and a denser $\Gamma$-centered Monkhorst--Pack $k$-point mesh of $6\times6\times6$ are employed to converge the force constants.
Following Ref.~\cite{grabowski2007ab}, the additional support-grid option (flag \textsc{addgrid}) is used to evaluate the augmentation charges, resulting in an augmentation grid of $384\times384\times384$.
Supplemental convergence tests are presented in Section~\ref{SSec_Conv_tests} of the Supplementary Materials.
For calculations at an electronic temperature of 0~K, the electronic occupancies are treated using the first-order Methfessel--Paxton smearing~\cite{MP1989} with a width of $0.1$~eV.
For the thermal electronic contribution $F^{\mathrm{el}}$, electronic occupancies are evaluated using Fermi smearing at the corresponding electronic temperature~\cite{mermin1965thermal}.

To derive the equilibrium thermal expansion of the bulk structure $V_{\mathrm{eq, bulk}}(T)$, the static and vibrational contributions are evaluated to cover the relevant part of the $V$- and $T$-dependent free-energy surface. Specifically, $E^{\mathrm{0K}}$ is computed at 17 volumes and the dynamical matrices entering $F^{\mathrm{qh}}$ are obtained at 6 volumes, while thermodynamic integration at the MTP level is performed on a $6 \times 6$ $(V, T)$ grid. The full DFT vibrational free-energy surface is then obtained by upsampling on a $4 \times 6$ $(V, T)$ grid, from which the equilibrium thermal-expansion path $V_{\mathrm{eq, bulk}}(T)$ of B2 MoTa is determined. For computing the Gibbs energies of the defect structures (vac$_{\mathrm{Mo}}$, vac$_{\mathrm{Ta}}$, Mo$_{\mathrm{Ta}}$, Ta$_{\mathrm{Mo}}$), the same grid setup is employed up to the MTP-level as for the bulk. Subsequent steps, namely the direct upsampling to DFT accuracy for both vibrational and electronic free-energy contributions, are performed along the bulk thermal-expansion path $V_{\mathrm{eq, bulk}}(T)$, using results obtained at five temperatures. For consistency, the corresponding upsampling terms of the bulk reference are recomputed at the same temperatures. 
Further details about the sampling grids and the parameterizations of the free-energy surface are provided in Section~\ref{SSec_Comp_Para} of the Supplementary Materials.

A level-24 MTP with a cutoff radius of 5~\AA\  (labeled as {\tt MTP\_up} in the Supplementary Materials) is used for the thermodynamic integration and the follow-up direct upsampling. This MTP is fitted to energies and forces (excluding stresses) of molecular-dynamics snapshots generated at 1000, 2400, and 3000~K. The training set spans multiple sub-systems, including ordered B2 MoTa (bulk and point-defect configurations), disordered MoTa, and pure BCC Mo and Ta, with a total of 4731 configurations. The resulting training root-mean-squared errors (RMSEs) are 2.53~meV/atom in energies and 0.23~eV/\AA~in forces. The broad structural diversity enables stable sampling for both bulk and defect models and provides a reliable reference ensemble for free-energy perturbation. An additional level-20 MTP (labeled as {\tt MTP\_vib} in the Supplementary Materials) with a cutoff radius of 5~\AA\ is trained as a task-specific potential restricted to ordered B2 MoTa. This MTP is used only to analyze the local vibrational distributions in the vicinity of vacancies. 
Its training set contains 2007 configurations, and the corresponding training RMSEs of energies and forces are 1.26~meV/atom and 0.16~eV/\AA, respectively. 
Further details on the training of the MTPs are provided in Section~\ref{SSec_MTP_training} of the Supplementary Materials.

The vibrational distribution of the atoms surrounding $\mathrm{vac_{Mo}}$ and $\mathrm{vac_{Ta}}$ (Figure \ref{F_PairDistri}) is sampled using molecular-dynamics simulations based on {\tt MTP\_vib}.
A total of 50 independent simulations, each of 10~ps with different initial velocity seeds, are performed.
For each trajectory, the relative position vectors between the vacancy site and its first- and second-nearest-neighbor (1st-NN and 2nd-NN) atoms are recorded. 
Symmetry-equivalent configurations are mapped onto a single representative orientation to enhance statistical sampling, yielding approximately $8\times10^5$ sampled configurations. 
The local density of sampled configurations is then evaluated for each trajectory point by counting the number of neighboring configurations within a spherical cutoff radius of 0.1~\AA~using OVITO~\cite{stukowski2009visualization}.
This procedure provides a quantitative measure of the local occupation of the vibrational phase space. 

%%%%%%%%%%%%%%%%%%%%%%%%%%%%%%%%%%%%%%%%%%%%%%%%%%%%%%%
%%%%%%%%%%%%%%%%%%%%%%%%%%%%%%%%%%%%%%%%%%%%%%%%%%%%%%%
\section{Results and discussion}

\subsection{Zero-Kelvin formation energies}
\label{Sec_vac_0K}

\begin{figure}[!hb]
    \centering
\includegraphics[width=0.48\textwidth]{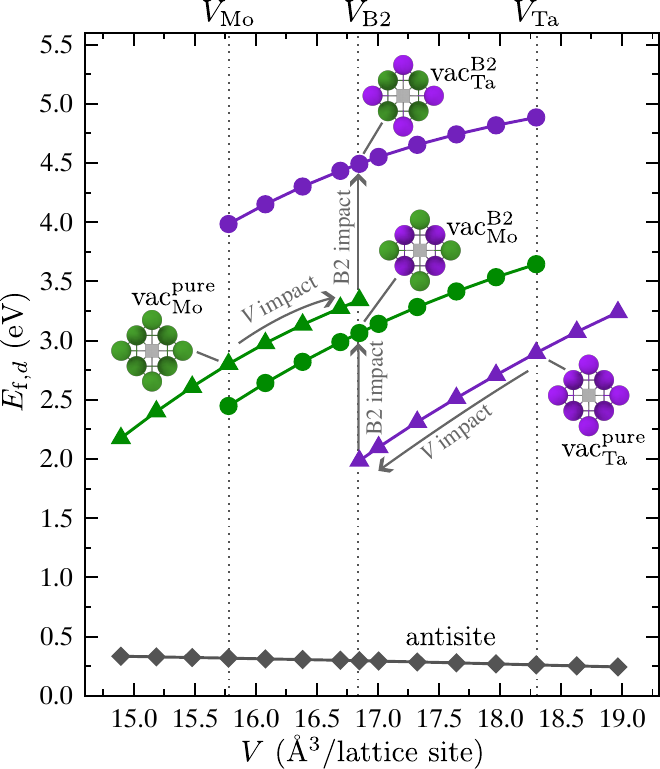}
\caption{\textbf{Zero-Kelvin formation energies of point defects in B2 MoTa and elemental BCC Mo/Ta.}   
The atomistic insets characterize the projected 1st- and 2nd-nearest neighbors of the four types of analyzed vacancies, with the gray cube at the center marking the vacancy site.
The indigo and green colors represent Ta and Mo atoms, respectively. 
The vertical dotted lines indicate the 0~K equilibrium volumes of the perfect crystal of pure Mo ($V_\mathrm{Mo}$), pure Ta ($V_\mathrm{Ta}$), and B2 MoTa ($V_\mathrm{B2}$). The ``B2 impact'' arrows correspond to the chemical environment change introduced by forming the B2 compound.
}
\label{F_vac_E_0K}
\end{figure}

Figure~\ref{F_vac_E_0K} shows the 0~K formation energies of the considered point defects as a function of volume. A clear difference can be seen between the antisite and the vacancy formation energies. The vacancy formation energies are significantly higher than those of antisites and increase strongly with volume. Antisite formation energies have a weak volume dependence and are about one order of magnitude lower than the vacancy formation energies. This finding is fully consistent with the determination of antisite pairs as the dominant point-defect combination (see Section~\ref{SSec_Dominant_defects} in the Supplementary Materials).

Another salient feature revealed by Figure~\ref{F_vac_E_0K} is that the two types of B2 vacancies exhibit a pronounced separation in their energies of formation (difference between the green and purple filled circles), e.g., 1.43~eV at the equilibrium B2 volume, $V_{\mathrm{B2}}$. 
To clarify the physical origin of the separation, the vacancy formation energies of elemental BCC Mo and Ta are also shown over the volume windows around their respective equilibrium states (filled triangles). 
As highlighted by the atomistic insets, chemical order imposes distinct chemistries in 1st-NNs: A vacancy on the Ta sublattice in B2 MoTa ($\mathrm{\mathrm{vac^{B2}_{Ta}}}$) is Mo-coordinated, similarly as a vacancy in elemental Mo ($\mathrm{vac^{pure}_{Mo}}$); vice versa, a vacancy on the Mo sublattice in B2 MoTa ($\mathrm{vac^{B2}_{Mo}}$) is Ta-coordinated, similarly as a vacancy in elemental Ta ($\mathrm{vac^{pure}_{Ta}}$).

Over the sampled volume windows, all four vacancy formation energies increase monotonically with $V$. 
Moreover, near their own equilibrium volumes, pure Mo and pure Ta show comparable vacancy formation energies. 
On this basis, the B2 vacancy splitting can be interpreted using the two-step construction indicated by the arrows in Figure~\ref{F_vac_E_0K}.

The arrows labeled as ``$V$ impact'' indicate how the two elemental references are brought to the common volume $V_{\mathrm{B2}}$ from their respective equilibrium volumes: The Mo-coordinated reference increases (effective expansion), whereas the Ta-coordinated reference decreases (effective compression). 
These opposite, volume-induced changes yield a large separation of 1.36 eV for two elemental systems at $V_{\mathrm{B2}}$.
The vertical arrows labeled as ``B2 impact'' denote the remaining host-dependent offset between the elemental references and the corresponding B2 vacancies (i.e., $\mathrm{vac^{pure}_{Mo}}\!\to \mathrm{vac^{B2}_{Ta}}$ and $\mathrm{vac^{pure}_{Ta}}\!\to \mathrm{vac^{B2}_{Mo}}$). 
This offset can be attributed to the extra energetic penalty of forming a vacancy in the chemically ordered B2 Mo--Ta environment and is nearly identical for the two branches, acting as an almost rigid upward shift.
Accordingly, the B2 vacancy splitting is dominated by the ``$V$ impact''.

%%%%%%%%%%%%%%%%%%%%%%%%%%%%%%%%%%%%%%%%%%%%%%%%%%%%%%%%%%%%
%%%%%%%%%%%%%%%%%%%%%%%%%%%%%%%%%%%%%%%%%%%%%%%%%%%%%%%%%%%%
\subsection{\texorpdfstring{\textit{T}}{Lg}-dependent chemical potentials}
\label{Sec_mu_T}

\begin{figure}[!tb]
    \centering
    \includegraphics[width=0.4\textwidth]{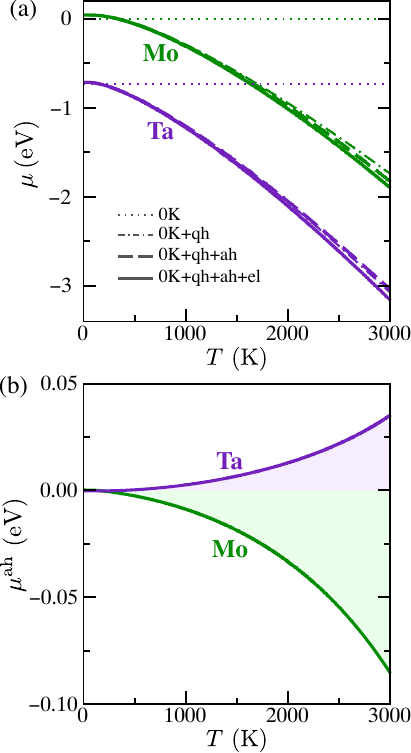}
\caption{\textbf{Chemical potentials in B2~MoTa.}
(a)~Chemical potentials of Mo and Ta, $\mu_{\mathrm{Mo}}(T)$ and $\mu_{\mathrm{Ta}}(T)$, referenced to $\mu_{\mathrm{Mo}}(0\,{\mathrm{K}})$, with the latter excluding zero-point vibrations.
The dotted line is the 0~K reference; the dotted-dashed, dashed, and solid lines progressively include quasiharmonic (qh), anharmonic (ah), and electronic (el) contributions.
(b)~Anharmonic contribution to the chemical potential, $\mu^{\mathrm{ah}}(T)$, for Mo and Ta.
}
\label{F_mu_T}
\end{figure}

Figure~\ref{F_mu_T}(a) shows the temperature-dependent chemical potentials $\mu_{\mathrm{Mo}}(T)$ and $\mu_{\mathrm{Ta}}(T)$, obtained by successively including thermal excitations: quasiharmonic, explicit anharmonic, and electronic. 
The offset between the Mo and Ta curves is related to the difference in cohesive energies between these elements.
Upon including thermal excitations, the chemical potentials decrease monotonically with increasing temperature, following a trend quantitatively similar to the Gibbs energies of pure Mo and Ta~\cite{forslund2023thermodynamic}.
The results obtained with different contributions indicate that the dominant temperature dependence originates from quasiharmonic vibrations, whereas explicit anharmonicity and thermal electronic excitations introduce smaller corrections.

Although the explicit anharmonic and electronic contributions are comparatively small, they can affect phase stabilities and derived thermodynamic quantities. In particular, the explicit anharmonic contribution to the chemical potential, i.e., $\mu^{\mathrm{ah}}$, exhibits opposite temperature trends for Mo and Ta in B2 MoTa, as shown in Figure~\ref{F_mu_T}(b).
For Mo, $\mu^{\mathrm{ah}}(T)$ decreases with temperature, whereas it increases for Ta.
This opposite trend in the anharmonic contributions is consistent with the group-resolved trends reported for unary BCC refractory metals, where Mo and W (group-VI elements) exhibit negative anharmonic contributions, while V, Nb, and Ta (group-V elements) exhibit positive ones~\cite{jung2023high, srinivasan2023anharmonicity, forslund2023thermodynamic}.
A quantitative comparison of $\mu^{\mathrm{ah}}(T)$ in B2 MoTa with the elemental chemical potentials is not pursued here, because the anharmonic contributions are strongly volume dependent and the thermal-expansion paths of B2~MoTa and the elemental systems differ substantially.

%%%%%%%%%%%%%%%%%%%%%%%%%%%%%%%%%%%%%%%%%%%%%%%%%%%%%%%%%%%%
%%%%%%%%%%%%%%%%%%%%%%%%%%%%%%%%%%%%%%%%%%%%%%%%%%%%%%%%%%%%
\subsection{Gibbs energy of antisite formation}
\label{Sec_anti_G}

\begin{figure}[!tb]
    \centering
    \includegraphics[width=0.41\textwidth]{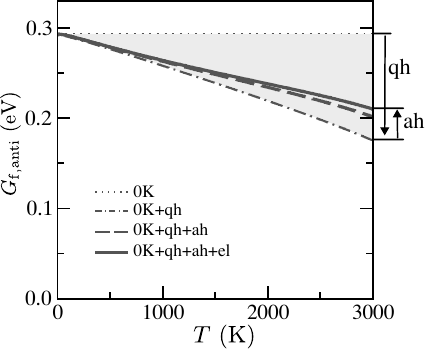}
\caption{\textbf{Gibbs energy of formation of antisites in B2~MoTa.}
The dotted line is the 0~K reference; the dotted-dashed, dashed, and solid lines progressively include quasiharmonic, anharmonic, and electronic contributions.
}
    \label{F_anti_T_G}
\end{figure}

The temperature-dependent Gibbs energy of formation of antisites in B2~MoTa is shown in Figure~\ref{F_anti_T_G}.
Across the considered temperature range, $\Ganti(T)$  gradually decreases from about $0.3$ to $0.2$~eV at 3000~K, i.e., a reduction of 30\%.
Specifically, the quasiharmonic contribution, highlighted by the gray-shaded region, dominates the temperature dependence of $\Ganti(T)$, while the anharmonic and electronic contributions partially compensate this reduction.

The overall small magnitude of $\Ganti(T)$ supports that antisite pairs remain the dominant point-defect combination in B2~MoTa (cf. Table~\ref{STab_point_combination_conc} in the Supplementary Materials). In other stoichiometric B2 compounds, formation energies of antisites can be substantially larger~\cite{leitner2015thermodynamics}. For example, in the triple-defect-dominant B2 NiAl compound, the 0~K antisite formation energies were reported to be in the range of $0.7$--$2.9$~eV~\cite{meyer1999atomic, leitner2015thermodynamics}. Speculating about a similar impact of the quasiharmonic contribution on a relative scale as observed here for B2 MoTa, a relevant influence on the thermodynamic balance between competing point-defect mechanisms can be expected in B2 NiAl.
A systematic investigation of the temperature-dependent antisite formation Gibbs energies in other B2 compounds would therefore be of considerable interest.

\begin{figure*}[!htb]
    \centering
    \includegraphics[width=0.9\textwidth]{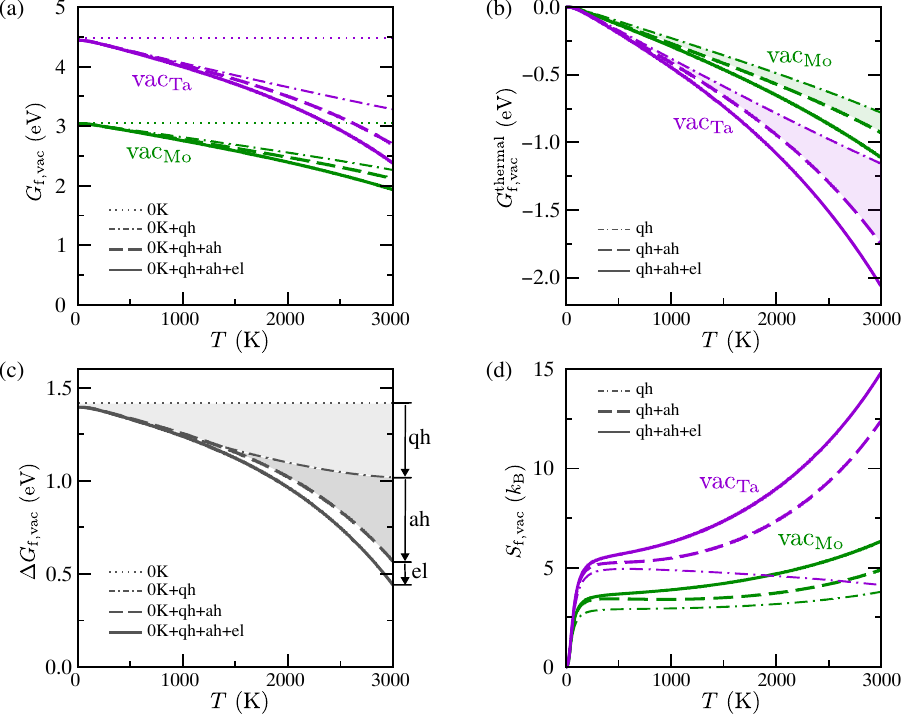}
    \caption{\textbf{Finite-temperature energetics of vacancies in B2~MoTa.}
    (a) Gibbs energies of vacancy formation, $G_{\mathrm{f, vac}}(T)$, on the Mo sublattice, $\mathrm{vac_{Mo}}$, and on the Ta sublattice, $\mathrm{vac_{Ta}}$.
    (b) Thermal contributions to the Gibbs energies of vacancy formation, $G_{\mathrm{f, vac}}^{\mathrm{thermal}}(T)$, referenced to the respective 0~K formation energies.
    (c) Difference of the Gibbs energies of vacancy formation, $\Delta G_{\mathrm{f, vac}}(T)=G_{\mathrm{f, vac_{Ta}}}(T)-G_{\mathrm{f, vac_{Mo}}}(T)$.
    (d) Entropies of vacancy formation, $S_{\mathrm{f, vac}}(T)$. The line styles indicate the inclusion of different thermal effects (0~K~=~static lattice, qh = quasiharmonic, ah = explicitly anharmonic, el = electronic).
    }
    \label{F_vac_T_G}
\end{figure*}

\subsection{Gibbs energies of vacancy formation}
\label{Sec_vac}
Figure~\ref{F_vac_T_G} summarizes the Gibbs energies of vacancy formation $G_{\mathrm{f, vac}}(T)$ in B2~MoTa, together with their thermal decomposition and the resulting entropy trends. 
As shown in Figure~\ref{F_vac_T_G}(a), $G_{\mathrm{f, vac}}(T)$ decreases with increasing temperature for both vacancy types, while the 0~K hierarchy $G_{\mathrm{f, vac_{Ta}}}(T)>G_{\mathrm{f, vac_{Mo}}}(T)$ is preserved over the entire temperature range. 
The thermal contributions are resolved in Figure~\ref{F_vac_T_G}(b). 
With inclusion of all thermal contributions, $G_{\mathrm{f, vac}}(T)$ decreases by $1.1$~eV for $\mathrm{vac_{Mo}}$ and by $2.1$~eV for $\mathrm{vac_{Ta}}$ between 0~K and 3000~K. 
Quasiharmonic vibrations dominate the temperature dependence for both vacancies.
Explicit anharmonicity contributes strongly beyond this baseline, and thermal electronic excitations provide an additional correction across the entire temperature range, becoming more apparent at high temperatures. 
The difference in the area of the shaded regions highlights the pronounced sublattice asymmetry of the anharmonic contribution. 
At 3000~K, the anharmonic contribution is $0.15$~eV for $\mathrm{vac_{Mo}}$ but reaches $0.62$~eV for $\mathrm{vac_{Ta}}$.

This temperature-dependent sublattice asymmetry is further highlighted in Figure~\ref{F_vac_T_G}(c) by the dependence of the difference $\Delta G_{\mathrm{f, vac}}(T)=G_{\mathrm{f, vac_{Ta}}}(T)-G_{\mathrm{f, vac_{Mo}}}(T)$. 
Including all thermal contributions (the solid curve), $\Delta G_{\mathrm{f, vac}}(T)$ decreases from an initial value of about $1.4$~eV at 0~K to about $0.4$~eV at 3000~K, clarifying that thermal contributions substantially narrow the energetic separation between $\mathrm{vac_{Ta}}$ and $\mathrm{vac_{Mo}}$ while preserving their ordering. 
The stepwise addition of excitation mechanisms further shows that all thermal excitations contribute to reducing $\Delta G_{\mathrm{f, vac}}$. 
In particular, both the quasiharmonic and explicit anharmonic contributions reduce $\Delta G_{\mathrm{f, vac}}(T)$ strongly (both by around $0.4$~eV at 3000~K). 
Their distinct physical origin is elucidated in Section~\ref{Sec_dG_vac}.
Thermal electronic excitations introduce a smaller additional change that becomes noticeable above 1500~K.

The entropies of vacancy formation are shown in Figure~\ref{F_vac_T_G}(d). 
As defined in Equation~\eqref{EQ_entropy}, the negative slope of $G_{\mathrm{f, vac}}(T)$ implies a positive entropy of formation. 
The nearly linear trends for both vacancy types when including quasiharmonic vibrations result in an approximately constant entropy of $\approx 4~k_{\mathrm{B}}$ over a broad temperature range. 
Upon including explicit anharmonicity (dashed curves), the entropy response becomes strongly vacancy dependent, reaching $12.5$~$k_{\mathrm{B}}$ for $\mathrm{vac_{Ta}}$, compared with $5.0$~$k_{\mathrm{B}}$ for $\mathrm{vac_{Mo}}$ at 3000~K.
Thermal electronic excitations contribute an additional $\approx 2~k_{\mathrm{B}}$ for both vacancies at this temperature.

A comparison with prior studies places the magnitude of the thermal contributions obtained here in context.
The net thermal reductions obtained at 3000~K, namely of $1.1$~eV for $\mathrm{vac_{Mo}}$ and $2.1$~eV for $\mathrm{vac_{Ta}}$, are substantially larger than those typically reported for vacancy formation in FCC elemental metals~\cite{grabowski2009ab, glensk2014PRX, gong2018Nivac}, where thermal effects near melting lower the vacancy formation Gibbs energy by only $\sim 0.1$--$0.4$~eV.
Part of this contrasting behavior likely reflects the generally larger vacancy formation energies in refractory systems, for which the vibrational entropy contribution---and thus the thermal reduction of $G_{\mathrm{f, vac}}(T)$---tends to scale with the underlying formation energy. 
At the same time, the much higher melting temperatures of refractory systems enable probing of vacancy thermodynamics at significantly higher absolute temperatures, further enhancing vibrational effects.
Consistently, for refractory BCC elements such as Mo and W, reductions of the vacancy formation Gibbs energy on the order of $1$~eV have been reported close to melting~\cite{zhang2025ab, forslund2023thermodynamic}, comparable to the magnitude found here for $\mathrm{vac_{Mo}}$ in B2~MoTa (Figure~\ref{F_vac_T_G}).
To the best of our knowledge, a reduction in vacancy formation energy as large as $2.1$~eV, as observed for $\mathrm{vac_{Ta}}$, has not been reported to date.
In ordered compounds, such a large sublattice-asymmetric reduction warrants special attention because it can substantially bias vacancy partitioning between inequivalent sublattices.
In addition, the thermal electronic contributions at 3000~K reach $0.2$--$0.3$~eV, comparable to the value reported for BCC W~\cite{zhang2025ab}.

\subsection{Origin of the sublattice asymmetry}
\label{Sec_dG_vac}

The sublattice asymmetry observed in $\Delta G_{\mathrm{f, vac}}(T)$ in Figure~\ref{F_vac_T_G}(c) has different origins, depending on the type of the underlying free-energy contribution.
The zero-Kelvin origin has already been discussed in Section~\ref{Sec_vac_0K} and attributed primarily to the ``$V$ impact'' as elucidated in Figure~\ref{F_vac_E_0K}.
To resolve the origins of the quasiharmonic and anharmonic asymmetry, we decompose $\Delta G_{\mathrm{f, vac}}(T)$ following Equation~\eqref{EQ_PDE} as
\begin{equation}\label{EQ_vac_mu}
\Delta G_{\mathrm{f, vac}} (T) =  \Delta G_{\mathrm{vac}} (T) + \Delta \mu (T),
\end{equation}
where $\Delta G_{\mathrm{vac}} (T) = G_{\mathrm{vac_{Ta}}} (T) - G_{\mathrm{vac_{Mo}}} (T)$ denotes the vacancy-structure difference and $\Delta \mu (T) = \mu_{\mathrm{Ta}} (T) - \mu_{\mathrm{Mo}} (T)$ the chemical-potential difference.
Using Equations~\eqref{EQ_mu_Mo} and \eqref{EQ_mu_Ta}, the chemical-potential difference can be written as
\begin{equation}\label{EQ_anti_mu}
\Delta \mu (T) = \frac{1}{2}\Delta G_{\mathrm{anti}} (T),
\end{equation}
with $\Delta G_{\mathrm{anti}} (T)=G_{\mathrm{Ta_{Mo}}} (T) -G_{\mathrm{Mo_{Ta}}} (T)$.
We recall that the relation in Equation~\eqref{EQ_anti_mu} applies because of the stoichiometric constraint for B2 MoTa together with the dominance of the antisite-pair mechanism (see Section~\ref{Sec_Method_B2}).

\begin{figure}[!htb]
\centering
\includegraphics[width=0.45\textwidth]{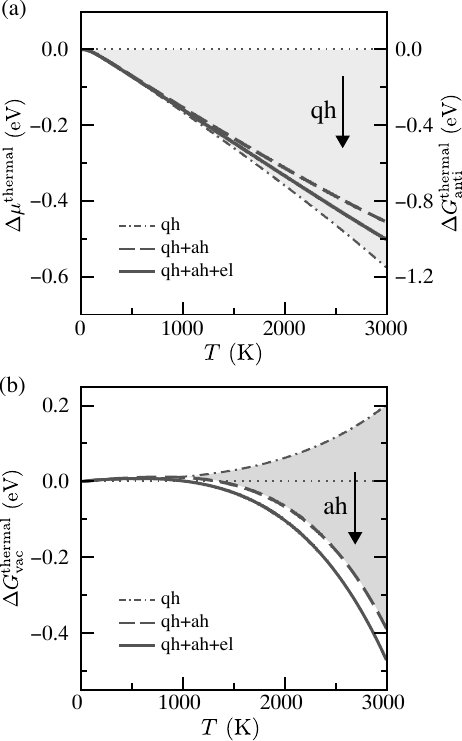}
\caption{\textbf{Thermal decomposition of chemical-potential and vacancy-structure terms. }
(a)~Thermal contribution to the chemical-potential difference, $\Delta \mu^{\mathrm{thermal}}(T)$, with $\,\Delta G_{\mathrm{anti}}^{\mathrm{thermal}} (T) = 2\,\Delta \mu^{\mathrm{thermal}}(T)$ on the secondary $y$-axis.
(b) Thermal contribution to the vacancy-structure difference, $\Delta G_{\mathrm{vac}}^{\mathrm{thermal}} (T)$.
The line styles indicate the stepwise inclusion of quasiharmonic (qh), explicit anharmonic (ah), and thermal electronic (el) contributions. 
The shading and the arrows highlight the dominant qh and ah contributions.
}
\label{F_dG_vac_thermal}
\end{figure}

Figure~\ref{F_dG_vac_thermal} shows the thermal contributions to the two terms in Equation~(\ref{EQ_vac_mu}) (defined relative to their zero-Kelvin values), with a detailed decomposition at 3000~K provided in Table~\ref{stab_deltaG_thermal_3000K} in the Supplementary Materials. 
Figure~\ref{F_dG_vac_thermal}(a) shows the chemical-potential difference $\Delta \mu^{\mathrm{thermal}}(T)$, with the related $\Delta G_{\mathrm{anti}}^{\mathrm{thermal}} (T)$ values emphasized on the secondary $y$-axis.
As indicated by the shaded region, the quasiharmonic contribution dominates the temperature dependence of $\Delta \mu^{\mathrm{thermal}}(T)$, while explicit anharmonicity has little impact. 
At 3000~K, the quasiharmonic contribution leads to a reduction in $\Delta \mu^{\mathrm{thermal}}(T)$ of $0.58$~eV.

In contrast, the quasiharmonic and anharmonic contributions to $\Delta G_{\mathrm{vac}}^{\mathrm{thermal}} (T)$ shown in Figure~\ref{F_dG_vac_thermal}(b) exhibit qualitatively different behaviors: The quasiharmonic contribution is positive and moderate over the entire temperature range, while the explicit anharmonic contribution is negative and becomes dominant at elevated temperatures, as highlighted by the gray shaded region. 
This pronounced anharmonicity drives $\Delta G_{\mathrm{vac}}^{\mathrm{thermal}}(T)$ to increasingly negative values, resulting at 3000~K in a reduction
% in $\Delta G_{\mathrm{f, vac}} (T)$ 
of $0.59$~eV.

The temperature dependence of the sublattice asymmetry in $\Delta G_{\mathrm{f, vac}}(T)$ shown in Figure~\ref{F_vac_T_G}(c) can thus be traced back to two distinct sources: The quasiharmonic contribution is primarily arising from the chemical-potential difference through the antisite-structure term $\Delta G_{\mathrm{anti}}^{\mathrm{thermal}} (T)$, whereas the anharmonic contribution is mainly arising from the vacancy-structure term $\Delta G_{\mathrm{vac}}^{\mathrm{thermal}} (T)$. 
At high temperatures, the quasiharmonic and anharmonic contributions associated with these two terms become comparable in magnitude and jointly determine the temperature dependence of $\Delta G_{\mathrm{f, vac}}(T)$. 
The electronic contributions to $\Delta G_{\mathrm{f, vac}}(T)$ are shared nearly equally between the vacancy- and antisite-structure terms.

The strong anharmonic contribution to $\Delta G_{\mathrm{vac}}(T)$ can be understood in terms of the local vibrational distributions around the vacancies.
Similar analyses were performed for vacancies in Cu, Ni, and W~\cite{glensk2014PRX, gong2018Nivac, zhang2025ab} and for planar defects in Ni$_3$Al~\cite{xu2023CSF}.
To this end, we train an additional dedicated MTP to sample the finite-temperature vibrational phase space of atoms surrounding the vacancy.
As shown in Section~\ref{SSec_MTP2} of the Supplementary Materials, this MTP closely reproduces the DFT-level temperature dependence of the vacancy formation Gibbs energies and also the anharmonic contribution to $\Delta G_{\mathrm{vac}}^{\mathrm{thermal}}(T)$, indicating that the local vibrational response relevant to vacancy formation is accurately captured.

\begin{figure}[!t]
    \centering
    \includegraphics[width=0.38\textwidth]{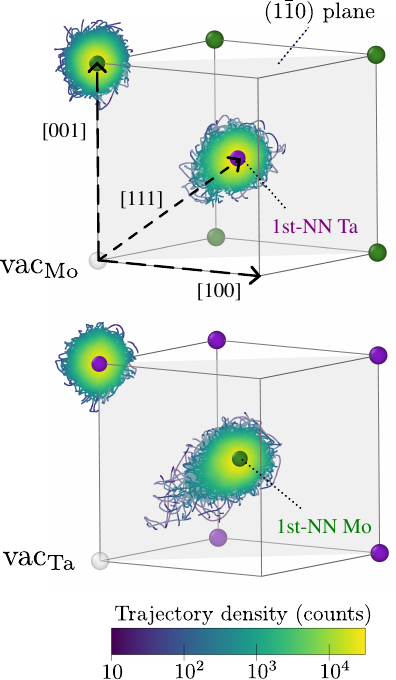}
    \caption{\textbf{Vibrational distributions of the atoms surrounding $\mathbf{vac_{Mo}}$ and $\mathbf{vac_{Ta}}$.}
    Sampling is performed at $a=3.29$~\AA\ and $T=2549$~K, corresponding to a representative state point along the thermal expansion path.
    The color of each trajectory point encodes the local, unnormalized trajectory-point density, defined as the number of sampled trajectory points within a cutoff radius of 0.1~\AA.
    The three-dimensional distribution is visualized by a slice on the $(1\bar{1}0)$ plane. Further details of the sampling are provided in Section~\ref{Sec_Method_Comp_para}.
    }
    \label{F_PairDistri}
\end{figure}

Figure~\ref{F_PairDistri} compares the resulting vibrational distributions of the atoms surrounding $\mathrm{vac_{Mo}}$ and $\mathrm{vac_{Ta}}$.
For $\mathrm{vac_{Mo}}$, the Ta atoms in the 1st-NN shell populate a compact distribution without pronounced directional tails, indicating a relatively confined local vibrational behavior.
In contrast, for $\mathrm{vac_{Ta}}$, the Mo atoms in the 1st-NN shell explore a substantially broader and more anisotropic distribution, with a clear elongation toward the vacancy center.
Such an enlarged and directional vibrational distribution reflects a weaker and anisotropic effective confinement of the local atomic vibrations around $\mathrm{vac_{Ta}}$, implying a larger anharmonic vibrational entropy and a stronger free-energy lowering than for $\mathrm{vac_{Mo}}$.
This real-space asymmetry thus provides the microscopic origin of the anharmonic reduction of $\Delta G_{\mathrm{vac}}^{\mathrm{thermal}}(T)$ and, consequently, of the temperature-dependent sublattice asymmetry in $\Delta G_{\mathrm{f, vac}}(T)$.

The present decomposition provides a physically transparent way to interpret the origin of temperature-dependent sublattice asymmetries of defect formation in ordered compounds. 
The resulting decomposition depends on the dominant defect mechanism. 
Given that thermal entropic contributions are material-specific, it is of particular interest to extend this decomposition analysis to other mechanisms in ordered compounds, such as the triple-defect mechanism in B2 NiAl~\cite{meyer1999atomic} and B2 PdAl~\cite{fu1995origin}, or the antisite-pair mechanism in L1$_2$ Ni$_3$Al~\cite{fu1997point}.
Specifically, in B2 NiAl, antisite formation energies at 0~K are significantly higher than those in B2 MoTa~\cite{meyer1999atomic, leitner2015thermodynamics}, resulting in a different balance between vacancy and antisite contributions.
This motivates examining how various vibrational contributions are partitioned among the terms contributing to the defect formation Gibbs energy.
%%%%%%%%%%%%%%%%%%%%%%%%%%%%%%%%%%%%%%%%%%%%%%%%%%%%%%%%%%%%
%%%%%%%%%%%%%%%%%%%%%%%%%%%%%%%%%%%%%%%%%%%%%%%%%%%%%%%%%%%%

\subsection{Antisite concentration}\label{Sec_anti_con}

\begin{figure}[!b]
    \centering
    \includegraphics[width=0.45\textwidth]{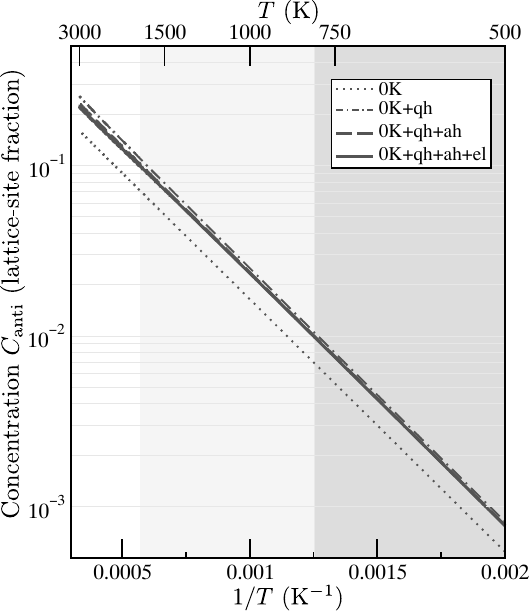}
\caption{
\textbf{Antisite concentration in B2 MoTa.} 
The line styles indicate the inclusion of different thermal effects in the underlying Gibbs energies of formation (cf.~Figure~\ref{F_anti_T_G}).
The darker gray region indicates the lower-temperature regime where B2 order is expected to be well developed.
The lighter gray region marks the literature-reported range of order--disorder transition temperatures, which originate from different theoretical approaches and exhibit a substantial scatter ($800\pm{200}$~K~\cite{blum2004mixed}, $1239$~K~\cite{zhang2022phase,wang2018computation}, $1770$~K~\cite{turchi2005application}).
}
    \label{F_anti_T_Con}
\end{figure}

Figure~\ref{F_anti_T_Con} shows the temperature dependence of the antisite concentration computed using Equation~\eqref{EQ_Concent} within the dilute-limit approximation.
For stoichiometric B2~MoTa, antisites occur on both sublattices with equal concentrations, i.e., $\Canti (T):=C_{\mathrm{Mo_{Ta}}}(T)\approx C_{\mathrm{Ta_{Mo}}}(T)$, to maintain stoichiometry. The total antisite concentration thus equals $2\Canti (T)$.

The equilibrium antisite concentration in Figure~\ref{F_anti_T_Con} reaches $7.5\times10^{-3}$ ($0.75\%$) already at 750~K and increases to $\sim 2.3\times10^{-2}$ ($2.3\%$) at 1000~K.
This trend reflects the small antisite formation Gibbs energies in Figure~\ref{F_anti_T_G}.
Using only the 0~K formation energy yields a slightly lower concentration than the curves that include finite-temperature contributions.
Moreover, the three curves with different thermal contributions are nearly indistinguishable, showing that additional anharmonic and electronic contributions have a negligible impact on the antisite concentration.

At low temperatures (below 750~K, dark gray region), where B2 order is expected to be well developed, antisite concentrations remain dilute, and the present dilute-limit description is therefore justified. 
With increasing temperature, the antisite concentration rises rapidly and reaches values of a few percent within the region of the theoretically predicted A2--B2 order--disorder transition temperatures. 
While we do not determine the transition temperature, the predicted antisite concentrations provide a useful reference for assessing the degree of sublattice mixing in this regime.
At high temperatures (above 1500~K), the antisite population becomes sufficiently large, and the dilute approximation underlying Equation~\eqref{EQ_Concent} breaks down.
Further discussion is provided in Section~\ref{Sec_Discussion}.

\subsection{Vacancy concentrations}

\begin{figure}[!hbt]
    \centering
    \includegraphics[width=0.45\textwidth]{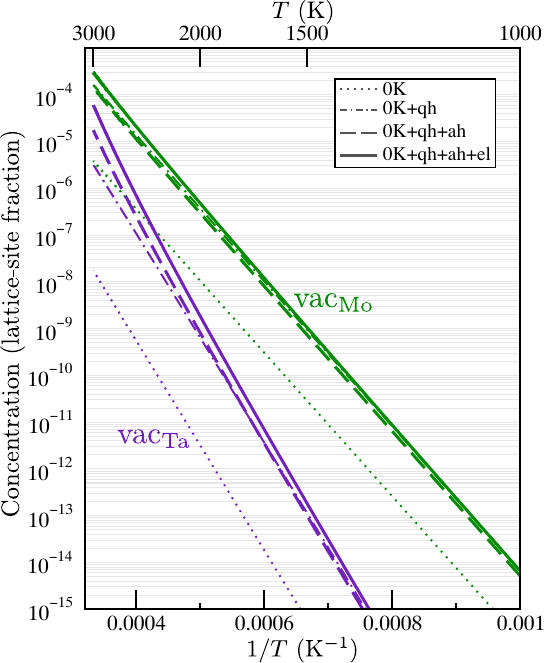}
\caption{\textbf{Vacancy concentrations in B2 MoTa.}
The concentrations of vacancies on the Mo and Ta sublattices, $\mathrm{vac_{Mo}}$ and $\mathrm{vac_{Ta}}$, respectively, are plotted as functions of inverse temperature. The line styles indicate the inclusion of different thermal effects in the underlying Gibbs energies of formation (cf.~Figure~\ref{F_vac_T_G}(a)).
% The shaded regions represent the effects of lattice vibrations.
}
    \label{F_vac_con}
\end{figure}

Figure~\ref{F_vac_con} shows the equilibrium vacancy concentrations in B2~MoTa as a function of inverse temperature $1/T$.
The vacancy concentrations derived from the 0~K formation energies exhibit a linear Arrhenius behavior, as shown by the straight dotted lines.
Including finite-temperature excitations, in particular anharmonicity, introduces non-linearity.
For both vacancy types, vibrational contributions dominate the increase of vacancy concentrations, whereas thermal electronic excitations provide a minor correction.
Notably, the impact of lattice vibrations is strongly sublattice dependent: $\mathrm{vac_{Ta}}$ exhibits a larger relative increase (exceeding three orders of magnitude) than $\mathrm{vac_{Mo}}$ (within two orders of magnitude), mirroring the larger vibrational free energy reduction of $\mathrm{vac_{Ta}}$ identified in Figure~\ref{F_vac_T_G}(b).

In the lower temperature range in Figure~\ref{F_vac_con}, the concentrations of the two types of vacancies differ by several orders of magnitude. However, at higher temperatures, $\mathrm{vac_{Ta}}$ becomes increasingly competitive.
While at all temperatures vacancies on the Mo sublattice site remain the dominant vacancy type, finite-temperature contributions raise $C_{\mathrm{vac_{Ta}}}(T)$ to about 20$\%$ of $C_{\mathrm{vac_{Mo}}}(T)$ at $3000$~K.
These concentration trends provide a direct population-level manifestation of the reduction in $\Delta G_{\mathrm{f, vac}}(T)$ at elevated temperatures (Figure~\ref{F_vac_T_G}(c)), and highlight the sublattice-dependent vacancy thermodynamics in ordered B2~MoTa.

\subsection{Beyond ideal B2: Impact of chemical disorder}\label{Sec_Discussion}

% \SD{probably to cite these papers of Anton den Ven? not all of them, but he might be a reviewer}
% \url{https://doi.org/10.1021/acs.chemmater.4c01903}
% \url{https://doi.org/10.1103/PhysRevB.94.094111}
% \url{https://doi.org/10.1016/j.actamat.2010.10.040}

The present analysis is formulated for stoichiometric B2 MoTa, in which chemical potentials and defect-formation Gibbs energies are internally consistent within the ordered reference state. 
When considering deviations from B2 order toward partially or fully disordered states, several additional thermodynamic aspects arise. 
Among the most relevant for the present defect-thermodynamic framework are: (i) the dilute-defect assumption for the vacancies, (ii) antisite populations that are intrinsically coupled to the degree of order, giving rise to an increased number of chemical environments and a different statistical treatment, and (iii) the elemental chemical-potential reference. 
Vacancies remain dilute up to the highest considered temperature, justifying the dilute-solution treatment, whereas antisite concentrations become appreciable toward the reported A2--B2 transition window.
In this regime, a coupled description of defect populations and the degree of order would be required for quantitative predictions in the partially disordered alloy~\cite{hu2004concentrated}, which lies beyond the present scope.

The third aspect concerns the choice of elemental chemical potentials, entering the defect-thermodynamic description as given in Equation~\eqref{EQ_PDE}.
We assess their robustness with respect to reductions in long-range order by comparing the present ordered B2 with the limit of a fully disordered A2 reference represented by a special-quasirandom structure (SQS)~\cite{zunger1990special} of the same system size. In the disordered limit, we evaluate chemical potentials within a DFT-based framework according to Refs.~\cite{zhang2022ab,luo2025determinants,dou2023first} (further details are provided in Section~\ref{SSec_mu_A2} of the Supplementary Materials).
Figure~\ref{F_mu_compare} shows that the chemical potentials obtained in the B2 reference remain close to the values in the A2 state.
The B2 chemical potentials fall within the statistical spread of the A2 distribution, indicating that the ordered reference state does not introduce an anomalous elemental bias.
Notably, the differences (below $\sim 0.1$~eV) are an order of magnitude smaller than the eV-scale thermal reductions identified for vacancy formation.
A rigorous finite-temperature treatment of the disordered state would require large-scale sampling and remains a subject for future work. 
Our present findings establish the B2 reference as a controlled thermodynamic baseline for vacancy formation and suggest that the main thermal trends remain relevant in regimes where B2-type local motifs persist as short-range order above the long-range order--disorder transition.

%\SD{and what is about composition dependence? Do the differences remain small for off-stoichiometric compositions?}
% \SD{is it possible to compare the vacancy spectra in A2 and B2 (in B2 just two delta functions)? this might indicate an impact of atom diffusion at A2--B2 transition, if averaged vacancy formation energy in A2 is very different to that of Va$_{\mathrm{Mo}}$. If not, diffusion mechanism, barriers, and jump correlations will determine (not assessed here).}

\begin{figure}[!tb]
    \centering
    \includegraphics[width=0.47\textwidth]{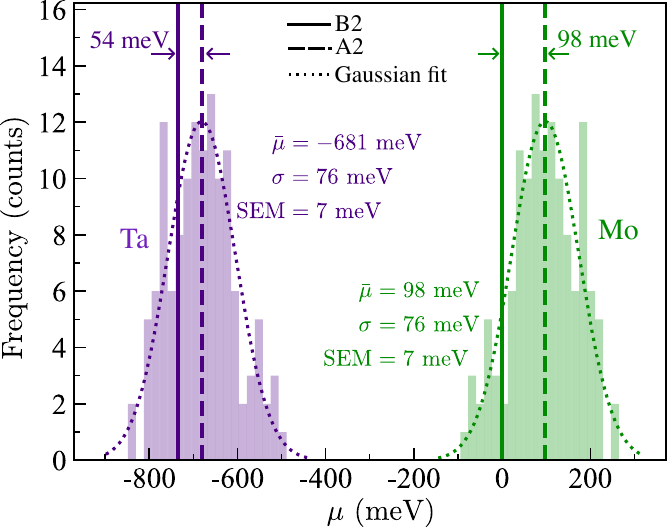}
\caption{\textbf{Zero-Kelvin chemical potentials in B2-ordered and A2-disordered MoTa.} 
The B2 chemical potentials are indicated by the solid vertical lines, purple for Ta and green for Mo. The distributions of A2 chemical potentials are shown by the light-colored histogram bars. The dotted curves denote Gaussian fits to the distributions. The corresponding mean values ($\bar{\mu}$), standard deviations ($\sigma$), and standard errors of the mean (SEM) are given in the figure. The mean values are additionally indicated by the dashed vertical lines.
All values are referenced to the B2 chemical potential of Mo.
    }
    \label{F_mu_compare}
\end{figure}

%%%%%%%%%%%%%%%%%%%%%%%%%%%%%%%%%%%%%%%%%%%%%%%%%%%%%%%%%%%%
%%%%%%%%%%%%%%%%%%%%%%%%%%%%%%%%%%%%%%%%%%%%%%%%%%%%%%%%%%%%
\section{Conclusion}\label{Sec_Conclusion}

A central result of this work is the pronounced sublattice asymmetry in the temperature dependence of the Gibbs energies of vacancy formation in B2~MoTa. From 0~K to 3000~K, the Gibbs energy of formation of a Ta-sublattice vacancy decreases by 2.1~eV and of a Mo-sublattice vacancy by 1.1~eV, revealing a large thermal reduction, particularly for the Ta-sublattice vacancy.

This temperature-dependent sublattice asymmetry arises from two distinct sources. 
The quasiharmonic contribution arises mainly from the chemical-potential imbalance set by the two antisite structures.
In contrast, the explicit anharmonic contribution arises primarily from the vacancy structures themselves. 
At high temperatures, these two contributions become comparable in magnitude and jointly determine the reduction in the difference between the Gibbs energies of Ta and Mo vacancy formation.

The atomistic origin of the anharmonic asymmetry is a site-specific vibrational softening around the Ta-sublattice vacancy.
The Mo atoms surrounding it explore a larger and more anisotropic vibrational space than the Ta atoms surrounding a Mo-sublattice vacancy, with enhanced displacement amplitudes toward the vacancy site.
This expanded vibrational distribution reflects a weaker local confinement and results in an enhanced anharmonic entropy contribution for the Ta-sublattice vacancy. 

These asymmetric thermal effects directly affect the equilibrium vacancy populations. Thermal excitations strongly increase the vacancy concentrations and reduce the population difference between the two sublattices. 
Consequently, Ta-sublattice vacancies become increasingly competitive with Mo-sublattice vacancies at elevated temperatures, highlighting the importance of finite-temperature vibrational effects for vacancy partitioning in ordered compounds.
Since vacancies are the primary carriers of atomic transport, temperature-dependent changes in vacancy partitioning suggest that atomistic diffusion pathways in ordered compounds (typically also sublattice-dependent) may vary with temperature.

The present results establish a direct connection between local chemical environment, vibrational space, and defect thermodynamics in an ordered B2 alloy. 
Since B2-type ordering and A2/B2 coexistence are frequently reported in refractory alloys~\cite{zhuo2024review}, site-dependent thermal-vibrational contributions, including explicit anharmonicity, are expected to be relevant beyond the present B2~MoTa reference system.

%%%%%%%%%%%%%%%%%%%%%%%%%%%%%%%%%%%%%%%%

\section*{Data Availability}
The data for the computed thermal properties of B2 MoTa and the Gibbs energies of formation of point defects will be openly available on DaRUS once the manuscript gets accepted.

\section*{Acknowledgements}
We thank Andrei V. Ruban for providing the spcm program for generating the special quasi-random structures. We acknowledge the funding from the European Research Council (ERC) under the European Union’s Horizon Europe Research and Innovation Programme (Grant Agreement No. 101200433, project META-LEARN).
X.X. and X.Z. acknowledge the financial support from the German Research Foundation (DFG) with the research grant ZH 1218/1-1 (project number 509804947). 
F.K. acknowledges support by the Deutsche Forschungsgemeinschaft (DFG) - Heisenberg Programme under grant number 541649719. 
S.D. acknowledges the financial support from the German Research Foundation (DFG) with the research grant DI 1419/24-1 (project number 509804947). 
X.X. gratefully acknowledges the scientific support and HPC resources provided by the Erlangen National High Performance Computing Center (NHR@FAU) of the Friedrich-Alexander-Universität Erlangen-Nürnberg (FAU) under the NHR project a102cb. 
NHR funding is provided by federal and Bavarian state authorities. NHR@FAU hardware is partially funded by the German Research Foundation (DFG) -- 440719683.
X.X. also acknowledges the support by the state of Baden-Württemberg through bwHPC and the German Research Foundation (DFG) through grant no INST 40/575-1 FUGG (JUSTUS 2 cluster), and the national supercomputer Hawk at the High Performance Computing Center Stuttgart (HLRS) under the grant number H-Embrittlement/44239.

Funded by the European Union. Views and opinions expressed are however those of the author(s) only and do not necessarily reflect those of the European Union or the
European Research Council Executive Agency. Neither the European Union nor the granting authority can be held responsible for them.

Language editing tools have been used to assist with sentence refinement for certain sections of this manuscript. 
The authors have subsequently reviewed and edited all content as needed and take full responsibility for the final version of the publication.
The authors declare that they have no known competing financial interests or personal relationships that could have appeared to influence the work reported in this paper.
%%%%%%%%%%%%%%%%%%%%%%%%%%%%%%%%%%%%%%%%%%%%%%%%%%%%%%%%%%%%%%%%%%%%%%%%%%%%%%
\bibliographystyle{unsrt}
\bibliography{biblio.bib}

\end{sloppypar}
\end{document}

\maketitle

\tableofcontents

\renewcommand\arraystretch{1.2}
\section{Determination of the dominant point defects in B2 MoTa\label{SSec_Dominant_defects}}
\subsection{Fractions of point-defect combinations}
The following combinations of antisites and vacancies preserve the B2 stoichiometry:
\begin{align}
\label{Gfanti}
    \textrm{Antisite pair:} && G_{\rm f, antipair}(T) &\,=\, G_{\rm Ta_{Mo}}(T) + G_{\rm Mo_{Ta}}(T) - 2 \, G_{\rm perf}(T)\\[.15cm]
    \textrm{Vacancy pair:} && G_{\rm f, vacpair}(T) &\,=\, G_{\rm vac_{Mo}}(T) + G_{\rm vac_{Ta}}(T) - \tfrac{2N-2}{N} \, G_{\rm perf}(T)\\[.15cm]
    \textrm{Triple defect 1:} && G_{\rm f, tridef_1}(T) &\,=\, 2\,G_{\rm vac_{Mo}}(T) + G_{\rm Mo_{Ta}}(T) - \tfrac{3N-2}{N} \, G_{\rm perf}(T)\\[.15cm]
\label{Gftp}
    \textrm{Triple defect 2:} && G_{\rm f, tridef_2}(T) &\,=\, 2\,G_{\rm vac_{Ta}}(T) + G_{\rm Ta_{Mo}}(T) - \tfrac{3N-2}{N} \, G_{\rm perf}(T)
\end{align}
These Gibbs energies of formation of point-defect combinations are written in terms of $G_d$ and $G_\textrm{perf}$, i.e., the Gibbs energies of the supercell with and without the defect $d$ (with $d \in \{\rm Ta_{\rm Mo}, \rm Mo_{\rm Ta}, \rm vac_{\rm Mo}, \rm vac_{\rm Ta} \}$) as computed at the level of density functional theory (DFT) in the present study. The resulting fractions of the defect combinations and of the induced point defects are obtained from (cf.~Section~\ref{sec:derivation})
\begin{align}
\label{Canti1}
    \textrm{Antisite pair:} &&  X_{\rm antipair}(T) = \exp \left( -\frac{G_{\rm f,antipair}(T)}{2 k_{\rm B} T} \right) && \rightarrow\quad X_{\rm antipair} = X'_{\rm Ta_{Mo}}=X'_{\rm Mo_{Ta}} \\[.15cm]
    \label{Canti2}
    \textrm{Vacancy pair:} && X_{\rm vacpair}(T) = \exp \left( -\frac{G_{\rm f,vacpair}(T)}{2 k_{\rm B} T} \right) && \rightarrow\quad X_{\rm vacpair} = X'_{\rm vac_{Mo}}=X'_{\rm vac_{Ta}} \\[.15cm]
    \label{Ctp1}
    \textrm{Triple defect 1:} && X_{\rm tridef_1}(T) = 2^{-2/3} \exp \left( -\frac{G_{\rm f,tridef_1}(T)}{3 k_{\rm B} T} \right) && \rightarrow\quad X_{\rm tridef_1} = X''_{\rm Mo_{Ta}}= \frac{1}{2} X''_{\rm vac_{Mo}} \\[.15cm]
\label{Ctp2}
    \textrm{Triple defect 2:} && X_{\rm tridef_2}(T) =  2^{-2/3} \exp \left( -\frac{G_{\rm f,tridef_2}(T)}{3 k_{\rm B} T} \right) && \rightarrow\quad X_{\rm tridef_2} = X''_{\rm Ta_{Mo}}= \frac{1}{2} X''_{\rm vac_{Ta}}
\end{align}
The primes indicate the distinct origin of the vacancy and antisite sublattice fractions, arising either from the pair combinations or the triple defect combinations. Tables \ref{STab_point_combination} and \ref{STab_point_combination_conc} show values for the computed Gibbs energies of formation and the resulting point-defect sublattice fractions at representative temperatures. Antisites arising from the formation of antisite pairs are dominant over the considered temperature range in B2 MoTa. A discussion on the non-dilute concentrations of antisites at high temperatures is provided in Section~3.8 in the main text.

\vspace{.2cm}

% Gibbs energies: 0K+qh+ah+el
\begin{table}[h!]
\centering
\caption{Temperature-dependent Gibbs energies of formation of point-defect combinations.}
\label{STab_point_combination}
\begin{tabularx}{0.71\textwidth}{c@{\hspace{26pt}}cccc}
\hline
$T$ (K) &
$G_{\rm f, antipair}$ (eV) & 
$G_{\rm f, vacpair}$  (eV) & 
$G_{\rm f, tridef_1}$  (eV) & 
$G_{\rm f, tridef_2}$  (eV) \vspace{0.02cm} \\  \hline
$0$    & 0.59 & 7.49 & 6.39 & 9.18 \\
$1000$ & 0.53 & 6.74 & 5.78 & 8.24 \\
$2000$ & 0.48 & 5.73 & 5.01 & 6.93 \\
$3000$ & 0.42 & 4.25 & 4.04 & 4.88 \\
\hline
\end{tabularx}
\vspace{0.4cm}
\end{table}

\begin{table}[h!]
\centering
\caption{Sublattice fractions based on Equations~\eqref{Canti1} to \eqref{Ctp2} and the values from Table \ref{STab_point_combination}.}
\label{STab_point_combination_conc}
\begin{tabular}{c c c cc cc}
\toprule
 & \multicolumn{1}{c}{Antisite pair} 
 & \multicolumn{1}{c}{Vacancy pair} 
 & \multicolumn{2}{c}{Triple defect 1} 
 & \multicolumn{2}{c}{Triple defect 2} \\
\cmidrule(lr){2-2}\cmidrule(lr){3-3}\cmidrule(lr){4-5}\cmidrule(lr){6-7}
$T$ (K) 
& $X'_{\rm Ta_{Mo}}=X'_{\rm Mo_{Ta}}$ 
& $X'_{{\rm vac}_{\rm Mo}}=X'_{{\rm vac}_{\rm Ta}}$ 
& $X''_{{\rm vac}_{\rm Mo}}$ 
& $X''_{\rm Mo_{Ta}}$ 
& $X''_{{\rm vac}_{\rm Ta}}$ 
& $X''_{\rm Ta_{Mo}}$ \\
\midrule
1000 & $4.6\times10^{-2}$ & $1.0\times10^{-17}$ & $3.9\times10^{-10}$ & $1.9\times10^{-10}$ & $2.9\times10^{-14}$ & $1.4\times10^{-14}$ \\
2000 & $2.5\times10^{-1}$ & $6.0\times10^{-8}$  & $1.2\times10^{-4}$  & $6.2\times10^{-5}$  & $3.0\times10^{-6}$  & $1.5\times10^{-6}$ \\
3000 & $4.4\times10^{-1}$ & $2.7\times10^{-4}$  & $1.1\times10^{-2}$  & $5.5\times10^{-3}$  & $3.7\times10^{-3}$  & $1.9\times10^{-3}$ \\
\bottomrule
\end{tabular}
\vspace{0.8cm}
\end{table}

\newpage
\subsection{Derivation of the defect-combination fractions\label{sec:derivation}}
For the configurational entropy, we assume that the B2 crystal can be factorized into two independent sublattices, $\alpha$ and $\beta$, each containing $M=N/2$ sites. In the dilute, non-interacting limit, the configurational counting therefore factorizes as $W = W_\alpha W_\beta$, because the occupation statistics on sublattice $\alpha$ are independent of those on sublattice $\beta$. Here, $W$ denotes the number of configurational realizations (microstates) for a given defect population. For symmetric defect combinations, such as an antisite pair or a vacancy pair, one defect is created on each sublattice. In this case,
\begin{equation}
W_\alpha = \binom{M}{n},
\qquad
W_\beta = \binom{M}{n},
\end{equation}
so that
\begin{equation}
W = \binom{M}{n}^2,
\end{equation}
where $n$ counts the point defects of a certain type and thus equals the number of defect combinations. For asymmetric defect combinations, such as triple defect 1 or triple defect 2, two defects are created on one sublattice and one defect on the other. The defects on the same sublattice are likewise assumed to be non-interacting. For example,
\begin{equation}
W_\alpha = \binom{M}{2n},
\qquad
W_\beta = \binom{M}{n},
\end{equation}
and therefore
\begin{equation}
W = \binom{M}{2n}\binom{M}{n}.
\end{equation}
For the defect combinations considered here, this can be written compactly as
\begin{equation}
W = \prod_i \binom{M}{\nu_i n},
\end{equation}
with $\nu_i$ denoting the multiplicity of defects on sublattice $i$. In the dilute limit,
\begin{equation}
\binom{M}{\nu_in} = \frac{M!}{(\nu_in)!(M-\nu_in)!} \approx \frac{M^{\nu_in}}{(\nu_in)!}
\end{equation}
and thus
\begin{equation}
\ln W \approx \sum_i [\nu_i n \ln M - \ln (\nu_in)!].
\end{equation}
Using Stirling's approximation, we can further write
\begin{equation}
\ln W \approx \sum_i [\nu_i n \ln M - \nu_i n \ln (\nu_i n) + \nu_in].
\end{equation}
The change $\Delta G$ in Gibbs energy of the crystal due to the defects is
\begin{equation}
\Delta G = n G_{\rm f,dc}(T) - k_B T \ln W,
\end{equation}
where $G_{\rm f,dc}$ is the Gibbs energy of forming a defect combination. Minimization gives
\begin{equation}
\frac{\partial \Delta G}{\partial n} = G_{\rm f,dc}(T) + k_B T \sum_i \nu_i \ln \left( \nu_i\frac{n}{M} \right) \equiv 0.
\end{equation}
Introducing the fraction $X_{\rm dc}$ of the defect combination
\begin{equation}
\label{eq:fraction}
X_{\rm dc} = \frac{n}{M},
\end{equation}
we have
\begin{equation}
\prod_i (\nu_i X_{\rm dc})^{\nu_i} = \exp \left(- \frac{G_{\rm f,dc}(T)}{k_B T} \right).
\end{equation}
The normalization in Equation~\eqref{eq:fraction} is chosen for convenience, and the induced point-defect fractions follow directly from \(X_{\rm dc}\) and the multiplicities \(\nu_i\).
For the antisite pair and vacancy pair, we have $\nu_1=\nu_2=1$, and for the triple defect combinations $\nu_1=2$ and $\nu_2=1$, and vice versa, resulting in the sublattice fractions in Equations~\eqref{Canti1} to \eqref{Ctp2}.

\clearpage
%%%%%%%%%%%%%%%%%%%%%%%%%%%%%%%%%%%%%%%%%%%%%%%%%%%%%%%%%%%%%%%%%%%%%%%%%
%%%%%%%%%%%%%%%%%%%%%%%%%%%%%%%%%%%%%%%%%%%%%%%%%%%%%%%%%%%%%%%%%%%%%%%%%
\section{Decomposition of vacancy-formation Gibbs energies at 3000 K}\label{SSec_decompos_delta_G_f}

Table~\ref{stab_deltaG_thermal_3000K} lists the thermal contributions to the vacancy formation Gibbs energies and the corresponding decomposed quantities at $T=3000$~K.
For stoichiometric B2 MoTa, the relations $\Delta G_{\rm f,vac}^{\rm thermal} = \Delta G_{\rm vac}^{\rm thermal} + \Delta \mu^{\rm thermal}$ and $\Delta \mu^{\rm thermal} = 1/2\, \Delta G_{\mathrm{anti}}^{\mathrm{thermal}}$ hold for each contribution.
\begin{table}[H]
\centering
\caption{Thermal contributions to the Gibbs-energy components at $T=3000$~K relative to the 0~K reference (unit: eV).}
\label{stab_deltaG_thermal_3000K}
\begin{tabularx}{\textwidth}{cYYYYYY}
\hline \rule{0pt}{10pt}
Contributions &
$G_{\rm f,vac_{Ta}}^{\rm thermal}$ &
$G_{\rm f,vac_{Mo}}^{\rm thermal}$ &
$\Delta G_{\rm f,vac}^{\rm thermal}$  &
$\Delta G_{\rm vac}^{\rm thermal}$    &
$\Delta G_{\rm anti}^{\rm thermal}$   &
$\Delta \mu^{\rm thermal}$ \rule[-6pt]{0pt}{0pt} \\
\hline
qh            & $-1.17$ & $-0.79$ & $-0.38$ & $0.20$  & $-1.15$ & $-0.58$\\
qh+ah         & $-1.79$ & $-0.94$ & $-0.85$ & $-0.39$ & $-0.91$ & $-0.46$\\
qh+ah+el      & $-2.11$ & $-1.13$ & $-0.97$ & $-0.47$ & $-1.00$ & $-0.50$\\
\hline
\end{tabularx}
\end{table}

\clearpage
%%%%%%%%%%%%%%%%%%%%%%%%%%%%%%%%%%%%%%%%%%%%%%%%%%%%%%%%%%%%%%%%%%%%%%%%%
%%%%%%%%%%%%%%%%%%%%%%%%%%%%%%%%%%%%%%%%%%%%%%%%%%%%%%%%%%%%%%%%%%%%%%%%%
\section{Materials properties}\label{SSec_thermal_properties}

\subsection{Zero-Kelvin results for B2 MoTa and elemental systems}
\begin{table}[H]
\centering
\caption{Results for the 0~K properties obtained from DFT calculations and fitted to the Vinet equation of state. Listed are the equilibrium lattice constant $a_{\rm eq}$, the corresponding atomic volume $V_{\rm eq}$, the bulk modulus $B_{\rm eq}$, and its pressure derivative $B'_{\rm eq}$ for B2 MoTa and the elemental reference systems.}
\label{STab_0K_properties}
\small
\begin{threeparttable}
\begin{tabularx}{\textwidth}{YYYYY}
\hline
Material systems & 
$a_{\rm eq}$~(\AA) & 
$V_{\rm eq}$~(\AA$^3$/atom) &  
$B_{\rm eq}$ (GPa) & 
$B'_{\rm eq}$ \\ \hline 
MoTa    & 3.230 & 16.84 & 231 & 4.1 \\
Mo      & 3.160 & 15.78 & 261 & 4.2 \\
Ta      & 3.320 & 18.30 & 195 & 3.8 \\ \hline
Mo~\cite{forslund2023thermodynamic} & 3.16 & 15.8 & 259 & 4.2 \\
Ta~\cite{forslund2023thermodynamic} & 3.32 & 18.3 & 196 & 3.8 \\
\hline 
\end{tabularx}
\end{threeparttable}
\end{table}

\subsection{Thermal properties of bulk B2 MoTa}
\begin{figure}[H]
    \centering
    \includegraphics[width=0.9\textwidth]{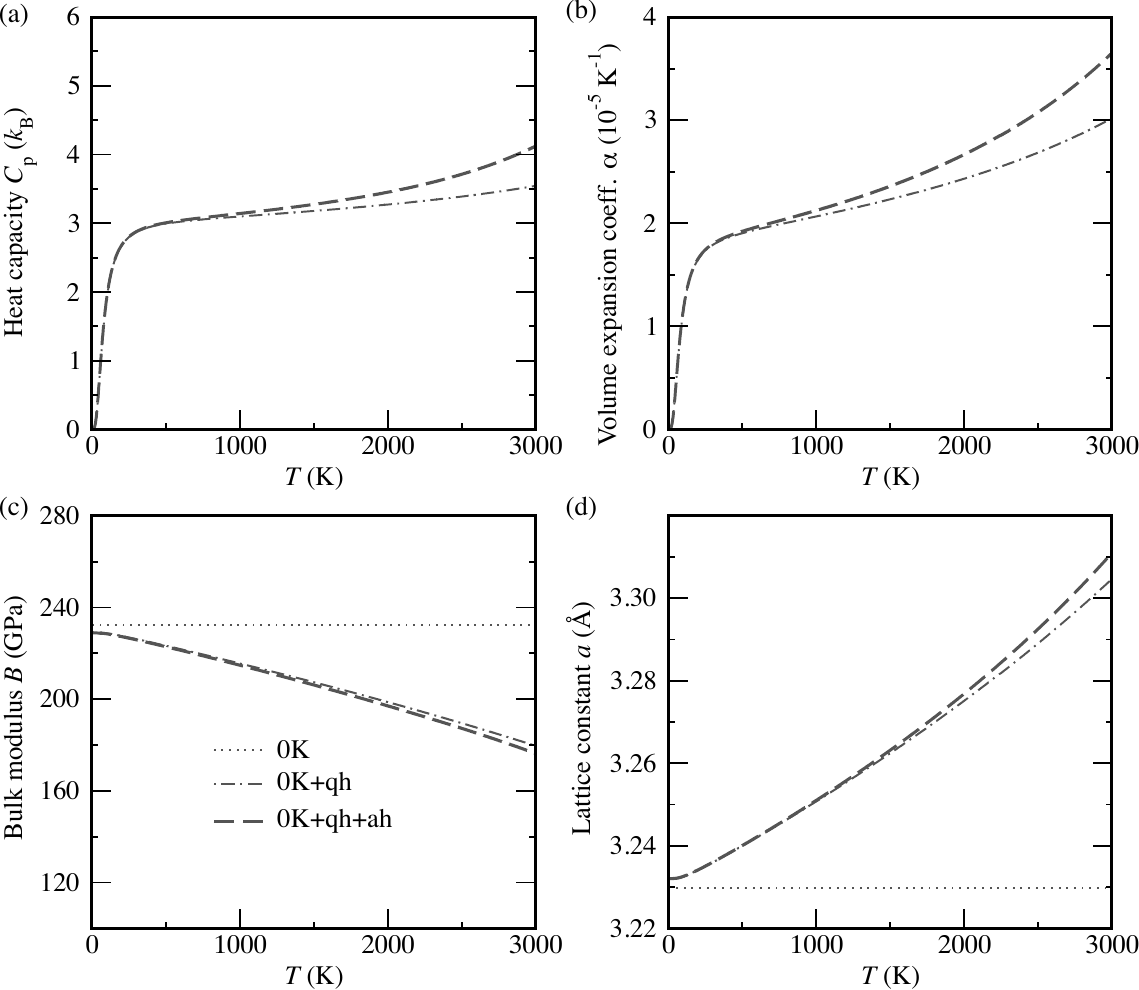}
\caption{Thermal properties of B2 MoTa with different thermal contributions.
(a) Heat capacity $C_p$; (b) volume expansion coefficient $\alpha$; (c) bulk modulus $B$; (d) lattice constant $a$.
}

    \label{sfig_TherPro}
\end{figure}

Figure~\ref{sfig_TherPro} shows the thermal properties of bulk B2 MoTa obtained from DFT, considering vibrational contributions.
As described in the Method section in the main text, thermal electronic contributions are estimated \textit{a posteriori} by evaluating $F^{\rm el}(T)$ along the equilibrium thermal-expansion path $V_{\mathrm{eq, bulk}}(T)$ obtained from the full vibrational free-energy surface (0K+qh+ah).
Thus, the electronic contributions to the thermal properties are not included in Figure~\ref{sfig_TherPro}.

\subsection{Electronic densities of states of B2 MoTa and elemental systems}

\begin{figure}[H]
    \centering
    \includegraphics[scale=0.8]{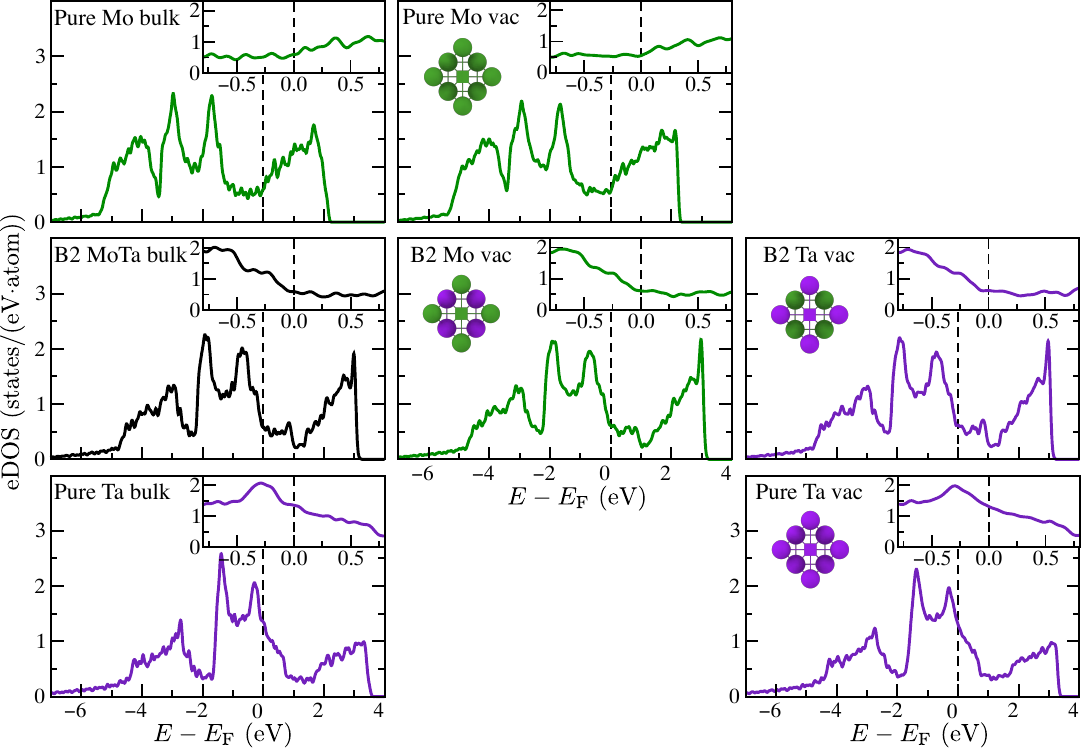}
\caption{Electronic density of states (eDOS) for unary reference crystals and B2 MoTa bulk/vacancy models as a function of energy referenced to the Fermi level, $E_{\mathrm{F}}$.
Insets show a zoom into the near-$E_{\rm F}$ region.}
    \label{sfig_DOSs}
    \vspace{0.5cm}
\end{figure}

\begin{figure}[H]
    \centering
    \includegraphics[scale=0.8]{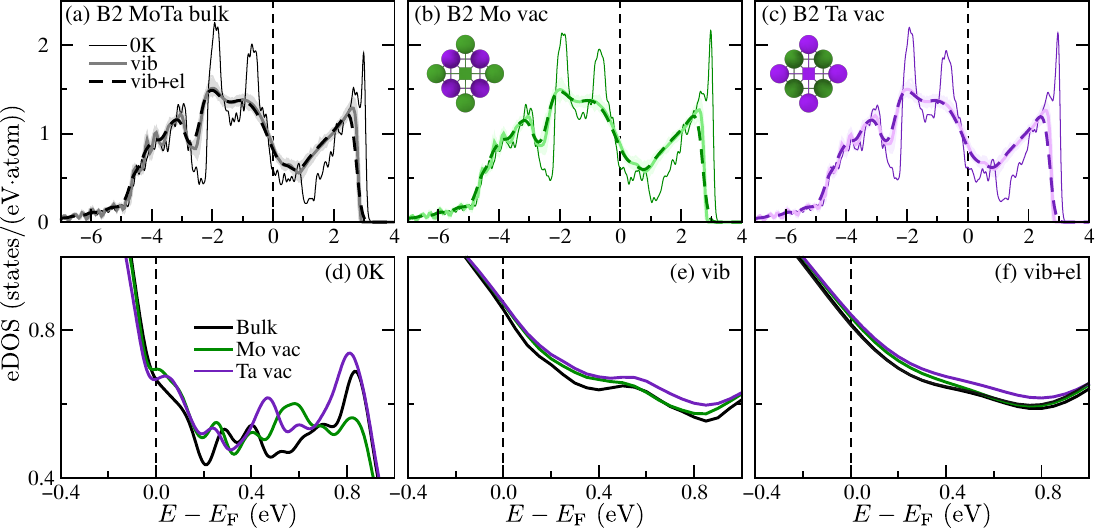}
\caption{Electronic density of states (eDOS) of bulk and vacancy models of B2 MoTa at different thermodynamic levels. 
(a-c) eDOS of B2~MoTa bulk, $\mathrm{vac_{Mo}}$, and $\mathrm{vac_{Ta}}$ evaluated for the 0~K reference (0K), the vibrationally expanded state (vib), and including the electronic contribution (vib+el).
(d-f) Zoom into the near-$E_{\rm F}$ region comparing bulk, $\mathrm{vac_{Mo}}$, and $\mathrm{vac_{Ta}}$ for the three levels (0K, vib, vib+el).}
    \label{sfig_DOSs-finite_T}
\end{figure}

\clearpage
%%%%%%%%%%%%%%%%%%%%%%%%%%%%%%%%%%%%%%%%%%%%%%%%%%%%%%%%%%%%%%%%%%%%%%%%%
%%%%%%%%%%%%%%%%%%%%%%%%%%%%%%%%%%%%%%%%%%%%%%%%%%%%%%%%%%%%%%%%%%%%%%%%%
\section{MTP training}\label{SSec_MTP_training}

Moment tensor potentials (MTPs)~\cite{shapeev2016moment,novikov2020mlip,gubaev2019accelerating} employed in this work are fitted with a loss function that includes energies and forces with weights $1.0$ and $0.01$~\AA$^2$, respectively, while stresses are not included.
Atomic interactions beyond a cutoff radius $R_{\rm cut}=5$~\AA~are neglected. 
Training configurations are generated in a subsystem-resolved manner, i.e., for a considered subsystem at a temperature $T$, a dedicated configuration pool is constructed using an active-learning workflow with an upgrade strategy~\cite{ou2024atomistic} from 12g to higher levels. 
Two target potentials---{\tt MTP\_up} and {\tt MTP\_vib}---are fitted for different purposes. 
{\tt MTP\_up} is trained on a broad selection from multiple subsystem pools to provide robust sampling for the free-energy workflow, whereas {\tt MTP\_vib} is trained as a target-specific potential for B2 MoTa. 
Table~\ref{STab_MTP-training} summarizes the subsystem pools used to construct the training sets and the resulting fitting statistics.

\begin{table}[H]
\centering
\caption{Components of the subsystem configuration pools and the resulting fitting statistics for the two target MTPs.
For each subsystem at a temperature $T$, $N_{\rm sel}$ configurations are selected to form the training set for {\tt MTP\_up} or {\tt MTP\_vib}, with a total size of $N_{\rm tot}$. 
The energy and force root-mean-square errors (RMSEs) relative to the DFT reference are also listed.}
\label{STab_MTP-training}
\small
\begin{threeparttable}
\begin{tabularx}{\textwidth}{l|l|llXX}
\hline \hline
\multirow{10}{*}{{\tt MTP\_up}} & 
\multirow{8}{*}{\makecell[c]{Training\\set}}  &
Sub-systems   &  
Configuration types & 
$T$ (K) & 
\makecell[l]{$N_{\rm sel}$} \\ \cline{3-6}
& & Pure Mo   &  bulk, vacancy                & 3000  & 1053 \\
& & Pure Ta   &  bulk, vacancy                & 3000  & 549  \\
& & Pure Ta   &  bulk, vacancy                & 1000  & 534  \\
& & B2 MoTa   &  bulk, vacancies, antisites   & 2400  & 539  \\
& & B2 MoTa   &  bulk, vacancies, antisites   & 1000  & 465  \\
& & SQS MoTa  &  bulk, vacancies              & 3000  & 1181 \\ 
& & SQS MoTa  &  \makecell[l]{bulk, vacancies, off-stoichiometric confs.} & 0    & 282 \\[0.1cm]\cline{2-6}
& \multirow{2}{*}{\makecell[c]{Training\\outputs}} &  \rule{0pt}{15pt} Level & $N_{\rm tot}$ & \makecell[l]{Energy RMSE \\ (meV/atom)} & \makecell[l]{Force RMSE \\ (eV/\AA)} \rule[-10pt]{0pt}{0pt} \\  \cline{3-6}
& & 24g   & 4731 & 2.53 & 0.23 \\
\hline \hline
\multicolumn{6}{c}{ } \vspace{-0.15cm}\\
\hline\hline
\multirow{5}{*}{{\tt MTP\_vib}} &
\multirow{3}{*}{\makecell[c]{Training\\set}} & Sub-systems   &  
Configuration types & $T$ (K) & \makecell[l]{$N_{\rm sel}$} \\ \cline{3-6}
& & B2 MoTa   &  bulk, vacancies, antisites   & 2400  & 1077  \\
& & B2 MoTa   &  bulk, vacancies, antisites   & 1000  & 930  \\[0.1cm] \cline{2-6}
& \multirow{2}{*}{\makecell[l]{Training\\outputs}} &  \rule{0pt}{15pt} Level & $N_{\rm tot}$ & \makecell[l]{Energy RMSE \\ (meV/atom)} & \makecell[l]{Force RMSE \\ (eV/\AA)} \rule[-10pt]{0pt}{0pt} \\  \cline{3-6}
& & 20g & 2007 & 1.26 & 0.16 \\
\hline \hline
\end{tabularx}
\end{threeparttable}
\end{table}

\clearpage
%%%%%%%%%%%%%%%%%%%%%%%%%%%%%%%%%%%%%%%%%%%%%%%%%%%%%%%%%%%%%%%%%%%%%%%%%
%%%%%%%%%%%%%%%%%%%%%%%%%%%%%%%%%%%%%%%%%%%%%%%%%%%%%%%%%%%%%%%%%%%%%%%%%
\section{Calculation parameters}\label{SSec_Comp_Para}

\subsection{Free-energy parametrization}
\label{SSSec_FES_para}

The anharmonic free energy $F^{\textrm{ah}}$, defined relative to the quasiharmonic reference, is evaluated via the direct-upsampling method~\cite{jung2023high} as
\begin{equation}\label{Seq_Fah}
    F^{\textrm{ah}} = \Delta F^{\textrm{qh} \rightarrow \textrm{MTP}} + \Delta F^{\textrm{up}}.
\end{equation}
Here, the MTP-level anharmonic contribution $\Delta F^{\textrm{qh} \rightarrow \textrm{MTP}}$ is obtained from thermodynamic integration as
\begin{equation}\label{Seq_integration}
    \Delta F^{\textrm{qh} \rightarrow \textrm{MTP}}
    = \int_0^1 {\textrm{d}}\lambda
    \left\langle E_{\textrm{MTP}} - E_{\textrm{qh}} \right\rangle_{\lambda},
\end{equation}
where the coupling parameter $\lambda$ interpolates between the quasiharmonic and the MTP potential-energy surface, and $\langle\cdots\rangle_\lambda$ denotes the corresponding ensemble average at a given $\lambda$. 
Specifically, $E_{\rm qh}$ denotes the harmonic reference potential constructed from the DFT force-constant matrix and $E_{\rm MTP}$ the potential energy predicted by the MTP for the same ionic configuration.
The upsampling correction $\Delta F^{\textrm{up}}$, evaluated at zero electronic temperature, is calculated by free-energy perturbation theory as
\begin{equation}\label{Seq_upsampling}
   \Delta F^{\textrm{up}} = -k_{\textrm{B}} T \ln \left\langle \exp \left(-\frac{E_{\textrm{DFT}} - E_{\textrm{MTP}}}{k_{\textrm{B}} T} \right) \right\rangle_{\textrm{MTP}},
\end{equation}
where $E_{\textrm{DFT}}$ denotes the DFT total energy evaluated for configurations sampled from the MTP ensemble.

Similarly, the upsampling term including the thermal electronic contribution $\Delta F^{\textrm{up+el}}$ is analogously computed as
\begin{equation}\label{Seq_F_up+el}
    \Delta F^{\textrm{up+el}} = -k_{\textrm{B}} T \ln \left\langle \exp \left(-\frac{F^{T_{\textrm{el}}}_{\textrm{DFT}} - E_{\textrm{MTP}}}{k_{\textrm{B}}T}\right)\right\rangle_{\textrm{MTP}},
\end{equation}
and the thermal electronic contribution $F^{\textrm{el}}$ is given by
\begin{equation}\label{Seq_Fel}
    F^{\textrm{el}} = \Delta F^{\textrm{up+el}} - \Delta F^{\textrm{up}}.
\end{equation}
In Equation~\eqref{Seq_F_up+el}, the electronic free energy $F^{T_{\textrm{el}}}_{\textrm{DFT}}$ for each configuration from the MTP ensemble is calculated at electronic temperature $T_{\textrm{el}}$ (set to the ionic temperature $T$) following the Mermin formalism~\cite{mermin1965thermal}.
Note that $F^{\textrm{el}}$ is thus evaluated on thermally displaced configurations, and therefore reflects the influence of thermal atomic vibrations on the thermal electronic free energy contribution.

\begin{table}[H]
\centering
\caption{Sampling and parametrization of the static, quasiharmonic, and
anharmonic free-energy contributions for all investigated configurations.}
\begin{tabular}{c c c c c c c}
\hline
 & bulk & vac$_{\rm Mo}$ & vac$_{\rm Ta}$ & Mo$_{\rm Ta}$ & Ta$_{\rm Mo}$ & Parametrization \\
\hline
$E^{\rm 0K}$ & 17 $V$ & 17 $V$ & 17 $V$ & 17 $V$ & 17 $V$ & Vinet-EOS \\
$F^{\rm qh}$ & 6 $V$ & 6 $V$ & 6 $V$ & 6 $V$ & 6 $V$ & $\{1,V,V^2,V^3\}$ \\
$\Delta F^{\rm qh \rightarrow MTP}$ 
& $6\times6$ $(V,T)$ 
& $6\times6$ $(V,T)$ 
& $6\times6$ $(V,T)$ 
& $6\times6$ $(V,T)$ 
& $6\times6$ $(V,T)$ 
& $\{1,V,T,V^2,VT,T^2\}$ \\
\hline
\end{tabular}
\label{tab:FES_common}
\end{table}

\begin{table}[H]
\centering
\caption{Sampling and parametrization of the bulk upsampling used to construct
the DFT-accuracy vibrational free-energy surface and determine the equilibrium
thermal-expansion path $V_{\rm eq,bulk}(T)$.}
\begin{tabular}{c c c}
\hline
 & bulk & Parametrization \\
\hline
$\Delta F^{\rm up}$ & $4\times6$ $(V,T)$ & $\{1,V,T,V^2,VT,T^2\}$ \\
\hline
\end{tabular}
\label{tab:FES_bulk_up}
\end{table}

\begin{table}[H]
\centering
\caption{Sampling and parametrization of the upsampling and thermal electronic
contributions evaluated for all configurations along the bulk equilibrium
thermal-expansion path $V_{\rm eq,bulk}(T)$.}
\begin{threeparttable}
\begin{tabularx}{0.85\textwidth}{lXXXXXl}
\hline
 & \tnote{\dag}bulk & vac$_{\rm Mo}$ & vac$_{\rm Ta}$ & Mo$_{\rm Ta}$ & Ta$_{\rm Mo}$ & Parametrization \\
\hline
$\Delta F^{\rm up}$ 
& 5 $T$ 
& 5 $T$ 
& 5 $T$ 
& 5 $T$ 
& 5 $T$ 
& $\{1,T,T^2\}$ \\
$F^{\rm el}$ 
& 5 $T$ 
& 5 $T$ 
& 5 $T$ 
& 5 $T$ 
& 5 $T$ 
& $\{T^2\}$ \\
\hline
\end{tabularx}

\vspace{0.2cm}
    \begin{tablenotes}
        \footnotesize
        \item[\dag] The bulk upsampling is recomputed on this path. This step is required in the present workflow because the evaluation of $F^{\rm el}$ requires DFT energies without thermal electronic excitations for the same snapshots. This treatment also ensures a consistent reference for the defect formation energies.

    \end{tablenotes}

\end{threeparttable}
\label{tab:FES_path}
\end{table}

\subsection{Detailed calculation parameters for the free energies}
\begin{table}[H]
\centering
\caption{Parameters for calculating 0~K and quasiharmonic contributions and the thermodynamic integration.}
\label{STab_para_DFT_0K_qh}
\small
\begin{threeparttable}
\begin{tabularx}{\textwidth}{cccccc}
\hline \hline
\multicolumn{6}{c}{0~K calculations} \\ \hline
Models  & \makecell[c]{\tnote{\dag}Lattice \\ constants $a$ (\AA)}  & $k$-points & kp$\cdot$atoms & \makecell[c]{Electronic convergence \\ criterion (eV)} & \makecell[c]{Force convergence \\ criterion (eV/\AA)} \\ \hline 
% Models  & \tnote{\dag}Lattice constants $a$ (\AA) & $k$-points & kp$\cdot$atoms & Electronic convergence (eV) & Force convergence (eV/\AA)\\ \hline 
bulk    &  3.10--3.40, $\Delta a = 0.02$             & $8\times8\times8$ & 55296 & $1\times 10^{-7}$ & $0.02$   \\
$\rm vac_{Mo}$   &  3.10--3.40, $\Delta a = 0.02$             & $8\times8\times8$ & 55296 & $1\times 10^{-7}$ & $0.02$   \\
$\rm vac_{Ta}$   &  3.10--3.40, $\Delta a = 0.02$             & $8\times8\times8$ & 55296 & $1\times 10^{-7}$ & $0.02$   \\
$\rm Mo_{Ta}$   &  3.10--3.40, $\Delta a = 0.02$             & $8\times8\times8$ & 55296 & $1\times 10^{-7}$ & $0.02$   \\
$\rm Ta_{Mo}$  &  3.10--3.40, $\Delta a = 0.02$             & $8\times8\times8$ & 55296 & $1\times 10^{-7}$ & $0.02$   \\
\hline \vspace{-0.3cm}\\
\multicolumn{6}{c}{Phonon calculations} \\ \hline
Models  & \makecell[c]{Lattice \\ constants $a$ (\AA)}  & $k$-points & kp$\cdot$atoms & \makecell[c]{Electronic convergence \\ criterion (eV)}  &  \\ \hline 
% Models & Lattice constants $a$ (\AA) & $k$-points & kp$\cdot$atoms & Electronic convergence (eV) \\ \hline 
bulk   & 3.22--3.32, $\Delta a = 0.02$             & $6\times6\times6$ & 27648 & $1\times 10^{-7}$   \\
$\rm vac_{Mo}$  & 3.22--3.32, $\Delta a = 0.02$             & $6\times6\times6$ & 27648 & $1\times 10^{-7}$   \\
$\rm vac_{Ta}$  & 3.22--3.32, $\Delta a = 0.02$             & $6\times6\times6$ & 27648 & $1\times 10^{-7}$   \\
$\rm Mo_{Ta}$  & 3.22--3.32, $\Delta a = 0.02$             & $6\times6\times6$ & 27648 & $1\times 10^{-7}$   \\
$\rm Ta_{Mo}$  & 3.22--3.32, $\Delta a = 0.02$             & $6\times6\times6$ & 27648 & $1\times 10^{-7}$   \\
\hline \vspace{-0.3cm}\\
\multicolumn{6}{c}{Thermodynamic integration} \\ \hline
Models& \makecell[c]{Lattice \\ constants $a$ (\AA)}  & \multicolumn{2}{c}{Temperature $T$ (K)} & \multicolumn{2}{c}{Coupling parameter $\lambda$} \\ \hline 
bulk  & 3.22--3.32, $\Delta a = 0.02$ & \multicolumn{2}{c}{500, 1000, 1500, 2000, 2400, 2700} & \multicolumn{2}{c}{\makecell[c]{0, 0.2, 0.4, 0.6, 0.7, 0.8, 0.85, 0.9}} \\
$\rm vac_{Mo}$ & 3.22--3.32, $\Delta a = 0.02$ & \multicolumn{2}{c}{500, 1000, 1500, 2000, 2400, 2700} & \multicolumn{2}{c}{\makecell[c]{0, 0.2, 0.4, 0.6, 0.7, 0.8, 0.85, 0.9}} \\
$\rm vac_{Ta}$ & 3.22--3.32, $\Delta a = 0.02$ & \multicolumn{2}{c}{500, 1000, 1500, 2000, 2400, 2700} & \multicolumn{2}{c}{\makecell[c]{0, 0.2, 0.4, 0.6, 0.7, 0.8, 0.85, 0.9}} \\
$\rm Mo_{Ta}$ & 3.22--3.32, $\Delta a = 0.02$ & \multicolumn{2}{c}{500, 1000, 1500, 2000, 2400, 2700} & \multicolumn{2}{c}{\makecell[c]{0, 0.2, 0.4, 0.6, 0.7, 0.8, 0.85, 0.9}} \\ 
$\rm Ta_{Mo}$ & 3.22--3.32, $\Delta a = 0.02$ & \multicolumn{2}{c}{500, 1000, 1500, 2000, 2400, 2700} & \multicolumn{2}{c}{\makecell[c]{0, 0.2, 0.4, 0.6, 0.7, 0.8, 0.85, 0.9}} \\ 
\hline \hline
\end{tabularx}
    \begin{tablenotes}
        \footnotesize
        \item[\dag] An additional lattice constant point at $a_{\rm eq} = 3.230$~\AA~for B2 MoTa is included for the 0~K calculations.

    \end{tablenotes}
\end{threeparttable}
\end{table}

The coupling parameter $\lambda$ is sampled from 0 to 0.9 to ensure consistency across the sampled defect models.
Sampling vacancy structures at $\lambda=1$ is less efficient because atoms can readily diffuse around the vacancy at elevated temperatures on the sampled time scale, as also discussed in Ref.~\cite{forslund2023thermodynamic}.
To avoid such events while maintaining a consistent protocol, all sampled $(V, T, \lambda)$ state points are fitted with the tangent function up to a maximum coupling parameter of $\lambda_{\max}=0.9$.

The thermodynamic integration is nevertheless performed over the full interval $\lambda\in[0,1]$. 
To validate this treatment, we perform explicit tests at representative $(V, T)$ points where sampling up to $\lambda=1$ remains stable (for bulk and antisite structures, and at lower temperatures for vacancy structures as well). 
Free-energy changes obtained from fits using the full $\lambda$ range are compared with those based on data up to $\lambda=0.9$ combined with extrapolation to $\lambda=1$. 
The differences are below $0.01$~meV/atom, confirming that the extrapolation does not affect the resulting anharmonic free energies.

\begin{table}[H]
\centering
\caption{Parameters for the direct upsampling calculations for the full free-energy surface for the bulk supercell. $N_\mathrm{conf}$ is the number of sampled configurations per $(V, T)$ grid point. The energy uncertainty is the maximum statistical uncertainty of the upsampling energy among the selected $(V, T)$ state points, typically at the largest $V$ and highest $T$ point. }
\label{STab_up_para_bulk}
\small
\begin{threeparttable}
\begin{tabularx}{\textwidth}{cccccc}
\hline 
\makecell[c]{Lattice \\ constants $a$ (\AA)} & Temperature $T$ (K) & $k$-points &  \makecell[c]{Electronic convergence \\ criterion (eV)} & $N_\mathrm{conf}$ & \makecell[c]{Energy uncertainty \\ (meV/atom)}\\ \hline 
3.24--3.30, $\Delta a = 0.02$ & 500--3000, $\Delta T = 500$ & $4\times4\times4$ & $1\times 10^{-5}$  & 48 & 0.25\\
\hline 
\end{tabularx}
\end{threeparttable}
\end{table}

\begin{table}[H]
\centering
\caption{Parameters for the direct upsampling calculations for all supercells along the thermal expansion path of the perfect crystal. $N_\mathrm{conf}$ is the number of sampled configurations per lattice constant. The energy uncertainty is the maximum statistical uncertainty of the upsampling energy among the selected state points. }
\label{STab_up_para_defects}
\small
\begin{threeparttable}
\begin{tabularx}{\textwidth}{YcYcYc}
\hline 
Models & \makecell[c]{Selected lattice \\ constants  $a$ (\AA)} &  $k$-points & \makecell[c]{Electronic convergence \\ criterion (eV)} & $N_\mathrm{conf}$ & \makecell[c]{Energy uncertainty \\ (meV/atom)}\\ \hline 
bulk & 3.25--3.29, $\Delta a = 0.01$ & $4\times4\times4$ & $1\times 10^{-5}$  & 84 & $0.14$\\
$\rm vac_{Mo}$ & 3.25--3.29, $\Delta a = 0.01$ & $4\times4\times4$ & $1\times 10^{-5}$  & 84 & $0.18$\\
$\rm vac_{Ta}$ & 3.25--3.29, $\Delta a = 0.01$ & $4\times4\times4$ & $1\times 10^{-5}$  & 84 & $0.24$\\
$\rm Mo_{Ta}$  & 3.25--3.29, $\Delta a = 0.01$ & $4\times4\times4$ & $1\times 10^{-5}$  & 84 & $0.14$\\
$\rm Ta_{Mo}$  & 3.25--3.29, $\Delta a = 0.01$ & $4\times4\times4$ & $1\times 10^{-5}$  & 84 & $0.13$\\
\hline 
\end{tabularx}
\end{threeparttable}
\end{table}

\begin{table}[H]
\centering
\caption{Parameters for calculating the electronic thermal contributions for all supercells along the thermal expansion path of the perfect crystal. $N_\mathrm{conf}$ is the number of sampled configurations for a selected lattice constant. The energy uncertainty is the maximum statistical uncertainty of the electronic thermal contribution among the selected state points.}
\label{STab_el_para_defects}
\small
\begin{threeparttable}
\begin{tabularx}{\textwidth}{YcYcYc}
\hline 
Models & \makecell[c]{Selected lattice \\ constants  $a$ (\AA)} &  $k$-points & \makecell[c]{Electronic convergence \\ criterion (eV)} & $N_\mathrm{conf}$ & \makecell[c]{Energy uncertainty \\ (meV/atom)}\\ \hline 
% Models & \makecell[c]{Selected lattice \\ constants  $a$ (\AA)} &  $k$-points & Electronic convergence (eV) & $N_\mathrm{conf}$ & Energy uncertainty (meV/atom) \\ \hline 
bulk & 3.25--3.29, $\Delta a = 0.01$ & $4\times4\times4$ & $1\times 10^{-5}$  & 48 & $0.11$\\
$\rm vac_{Mo}$ & 3.25--3.29, $\Delta a = 0.01$ & $4\times4\times4$ & $1\times 10^{-5}$  & 48 & $0.09$\\
$\rm vac_{Ta}$ & 3.25--3.29, $\Delta a = 0.01$ & $4\times4\times4$ & $1\times 10^{-5}$  & 48 & $0.13$\\
$\rm Mo_{Ta}$  & 3.25--3.29, $\Delta a = 0.01$ & $4\times4\times4$ & $1\times 10^{-5}$  & 48 & $0.11$\\
$\rm Ta_{Mo}$  & 3.25--3.29, $\Delta a = 0.01$ & $4\times4\times4$ & $1\times 10^{-5}$  & 48 & $0.11$\\
\hline 
\end{tabularx}
\end{threeparttable}
\end{table}

\subsection{Parameters for DFT calculations for elemental systems at 0~K}
\begin{table}[H]
\centering
\caption{Parameters for calculating 0~K formation energies in elemental systems.}
\label{STab_para_Pure}
\small
\begin{threeparttable}
\begin{tabularx}{\textwidth}{YcYYcc}
\hline
Models  & \makecell[c]{\tnote{\dag}Lattice \\ constants $a$ (\AA)}  & $k$-points & kp$\cdot$atoms & \makecell[c]{Electronic convergence \\ criterion (eV)} & \makecell[c]{Force convergence \\ criterion (eV/\AA)} \\ \hline 
Mo bulk      &  3.10--3.36, $\Delta a = 0.02$   & $8\times8\times8$ & 55296 & $1\times 10^{-7}$ & $0.02$   \\
Mo vacancy   &  3.10--3.36, $\Delta a = 0.02$   & $8\times8\times8$ & 55296 & $1\times 10^{-7}$ & $0.02$   \\
Ta bulk      &  3.18--3.40, $\Delta a = 0.02$   & $8\times8\times8$ & 55296 & $1\times 10^{-7}$ & $0.02$   \\
Ta vacancy   &  3.18--3.40, $\Delta a = 0.02$   & $8\times8\times8$ & 55296 & $1\times 10^{-7}$ & $0.02$   \\
\hline
\end{tabularx}
    \begin{tablenotes}
        \footnotesize
        \item[\dag] The equilibrium lattice constant of B2 MoTa ($a = 3.230$~\AA) is used as an additional sampling point for the elemental systems. 
    \end{tablenotes}
\end{threeparttable}
\end{table}

\subsection{Convergence tests}\label{SSec_Conv_tests}
\begin{figure}[H]
    \centering
    \includegraphics[width=\textwidth]{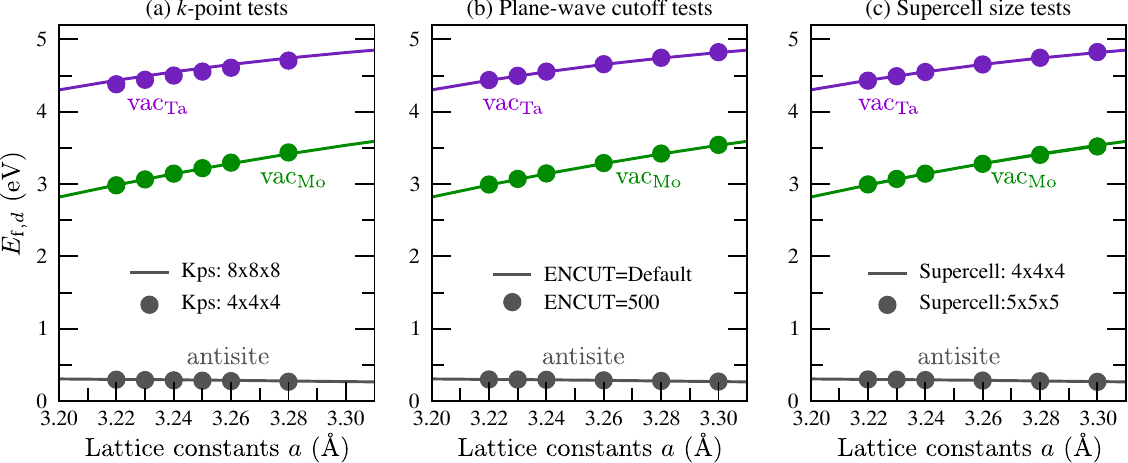}
\caption{Convergence tests for defect energetics.
Formation energies as a function of lattice constant $a$ obtained with modified numerical parameters relative to the standard settings (Table~\ref{STab_para_DFT_0K_qh}).
(a) $k$-point tests; (b) plane-wave cutoff tests; (c) supercell size tests.}

    \label{sfig_convergence_test}
\end{figure}

Figure~\ref{sfig_convergence_test} assesses the numerical convergence of the defect formation energies by repeating representative calculations with modified DFT settings relative to the standard protocol (Table~\ref{STab_para_DFT_0K_qh}).
All tests reproduce the same volume dependence and do not change the relative ordering between different defects.
The $k$-point convergence tests in Figure~\ref{sfig_convergence_test}(a) yield the largest deviation for $\mathrm{vac_{Ta}}$, with a maximum difference below $\sim 0.1$~eV over the sampled lattice-constant window.
In contrast, the changes induced by increasing the plane-wave cutoff in Figure~\ref{sfig_convergence_test}(b) and by enlarging the supercell in Figure~\ref{sfig_convergence_test}(c) are negligible for the present work.

\subsection{Parameters for calculating the free-energy surface with {\tt MTP\_vib}}
\begin{table}[H]
\centering
\caption{Parameters for calculating the free-energy surface with {\tt MTP\_vib}.}
\label{STab_para_MTP2}
\small
\begin{threeparttable}
\begin{tabularx}{0.95\textwidth}{cccc}
\hline \hline
\multicolumn{4}{c}{MTP-level 0~K and quasiharmonic calculations} \\ \hline
Models & Lattice constants $a$ (\AA) & Force convergence criterion (eV/\AA)  \\ \hline 
bulk   & 3.22--3.32, $\Delta a = 0.01$ & $1\times 10^{-10}$  \\
$\rm vac_{Mo}$  & 3.22--3.32, $\Delta a = 0.01$ & $1\times 10^{-10}$  \\
$\rm vac_{Ta}$  & 3.22--3.32, $\Delta a = 0.01$ & $1\times 10^{-10}$  \\
$\rm Mo_{Ta}$  & 3.22--3.32, $\Delta a = 0.01$ & $1\times 10^{-10}$  \\
$\rm Ta_{Mo}$  & 3.22--3.32, $\Delta a = 0.01$ & $1\times 10^{-10}$  \\
\hline \vspace{-0.3cm} \\
\multicolumn{4}{c}{MTP-level thermodynamic integration}  \\ \hline
Models& Lattice constants $a$ (\AA) & Temperature $T$ (K) & Coupling parameter $\lambda$ \\ \hline 
bulk  & 3.22--3.32, $\Delta a = 0.02$ & 300--2700, $\Delta T = 300$ & \makecell[c]{0, 0.2, 0.4, 0.6, 0.7, 0.8, 0.85, 0.9} \\
$\rm vac_{Mo}$ & 3.22--3.32, $\Delta a = 0.02$ & 300--2700, $\Delta T = 300$ & \makecell[c]{0, 0.2, 0.4, 0.6, 0.7, 0.8, 0.85, 0.9} \\
$\rm vac_{Ta}$ & 3.22--3.32, $\Delta a = 0.02$ & 300--2700, $\Delta T = 300$ & \makecell[c]{0, 0.2, 0.4, 0.6, 0.7, 0.8, 0.85, 0.9} \\
$\rm Mo_{Ta}$ & 3.22--3.32, $\Delta a = 0.02$ & 300--2700, $\Delta T = 300$ & \makecell[c]{0, 0.2, 0.4, 0.6, 0.7, 0.8, 0.85, 0.9} \\
$\rm Ta_{Mo}$ & 3.22--3.32, $\Delta a = 0.02$ & 300--2700, $\Delta T = 300$ & \makecell[c]{0, 0.2, 0.4, 0.6, 0.7, 0.8, 0.85, 0.9} \\
\hline \hline
\end{tabularx}
\end{threeparttable}
\end{table}

\clearpage
%%%%%%%%%%%%%%%%%%%%%%%%%%%%%%%%%%%%%%%%%%%%%%%%%%%%%%%%%%%%%%%%%%%%%%%%%
%%%%%%%%%%%%%%%%%%%%%%%%%%%%%%%%%%%%%%%%%%%%%%%%%%%%%%%%%%%%%%%%%%%%%%%%%
\section{Performance of {\tt MTP\_vib}}\label{SSec_MTP2}

\begin{figure}[H]
    \centering
    \includegraphics[width=0.8\textwidth]{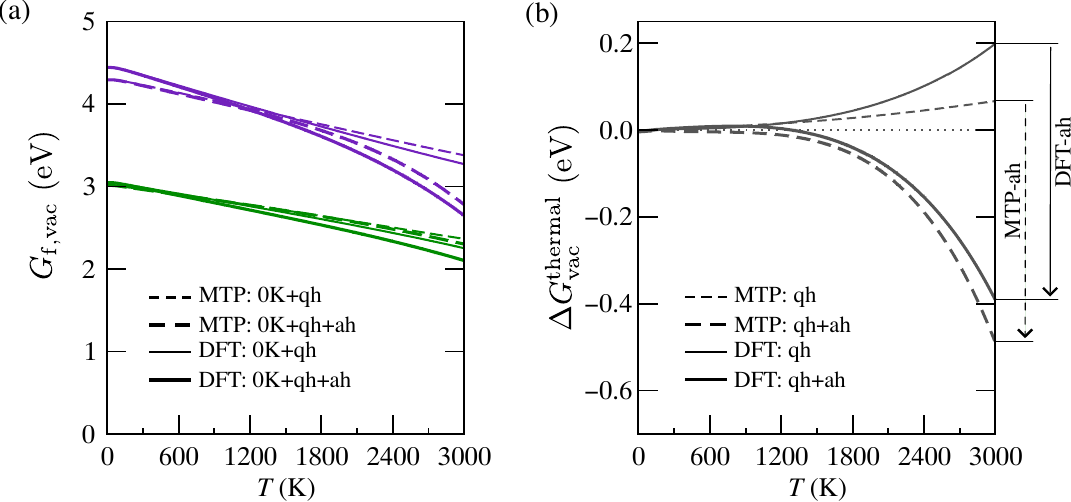}
\caption{Performance of {\tt MTP\_vib} for temperature-dependent vacancy formation energetics.
(a) Vacancy formation Gibbs energies $G_{\rm f,vac}(T)$ for $\mathrm{vac_{Mo}}$ and $\mathrm{vac_{Ta}}$ comparing {\tt MTP\_vib} and DFT at the 0K+qh and 0K+qh+ah levels.
(b) Thermal contribution of the vacancy-structure difference $\Delta G_{\rm vac}^{\rm thermal}$.}

    \label{sfig_DFT_VS_MTP}
\end{figure}

To accurately sample the vibrational phase space, a specific MTP for B2 MoTa is trained on a dataset that includes low-temperature configurations, as listed in Table~\ref{STab_MTP-training}.
Figure~\ref{sfig_DFT_VS_MTP} validates the {\tt MTP\_vib} potential by directly comparing the predicted vacancy formation Gibbs energies to the DFT results.
In panel~(a), the vacancy formation Gibbs energies $G_{\rm f, vac}(T)$ for $\mathrm{vac_{Mo}}$ and $\mathrm{vac_{Ta}}$ obtained from {\tt MTP\_vib} closely match the DFT results over the entire temperature range, for both the quasiharmonic level (0K+qh) and when explicit anharmonic effects are included (0K+qh+ah).
Panel~(b) further demonstrates the agreement in the vibrational contributions.

\clearpage
\section{Calculation of the chemical potentials in A2 MoTa}
\label{SSec_mu_A2}

The chemical potentials in A2 MoTa are computed at 0~K using the Widom-type insertion approach according to Refs.~\cite{zhang2022ab,luo2025determinants,dou2023first}. 
The random reference state of A2 MoTa is modeled using a special quasi-random structure (SQS)~\cite{zunger1990special}.
The total energy $E_{\mathrm{SQS}}$ of the reference SQS supercell satisfies the Euler relation as
\begin{equation}
E_{\mathrm{SQS}} = N_{\mathrm{Mo}} \mu_{\mathrm{Mo}} + N_{\mathrm{Ta}} \mu_{\mathrm{Ta}},
\end{equation}
where $N_{\rm Mo}$ and $N_{\rm Ta}$ are, respectively, the number of Mo and Ta atoms in the supercell.
A single-site substitution is performed by replacing one atom of species $i$ at site $s$ by species $j$, generating an off-stoichiometric configuration with the total energy $E_{i\rightarrow j}^{(s)}$ as
\begin{equation}
\Delta E_{i\rightarrow j}^{(s)}= E_{i\rightarrow j}^{(s)} - E_{\rm SQS},
\end{equation}
with $\Delta E_{i\rightarrow j}^{(s)}$ the corresponding substitution energy.
Within the Widom-type construction, this energy change is related to the chemical-potential difference as
\begin{equation}
\Delta E_{i\rightarrow j}^{(s)} = \mu_j^{(s)} - \mu_i^{(s)} .
\end{equation}
Solving the Euler relation for the reference SQS together with each site-specific substitution equation gives a site-resolved estimate $(\mu_{\mathrm{Mo}}^{(s)},\mu_{\mathrm{Ta}}^{(s)})$.

For the current SQS supercell containing 128 atoms, each single-site substitution yields a slightly different estimate. 
This results in 128 site-resolved estimates of the chemical potential, obtained from 64 Ta$\rightarrow$Mo and 64 Mo$\rightarrow$Ta substitutions (see Figure~8 in the main text). The chemical potentials can then be obtained as an average over the corresponding site-resolved values. In the present work, we use an arithmetic average, i.e., assigning equal statistical weight to each configuration (high-temperature limit), for comparison with the B2-ordered scenario.

The calculation parameters are provided in Table~\ref{STab_para_DFT_0K_A2}. 
 The equilibrated lattice constant of bulk A2 at 0~K is used for all configurations.
\begin{table}[H]
\centering
\caption{Parameters for calculating the A2 chemical potentials at 0~K, with $N_{\rm conf}$ the number of calculated configurations.}
\label{STab_para_DFT_0K_A2}
\small
\begin{threeparttable}
\begin{tabularx}{\textwidth}{YYYccccc}
\hline \hline
& Models  & $N_{\rm conf}$ & $k$-points & \makecell[c]{Electronic convergence \\ criterion (eV)} & \makecell[c]{Force convergence \\ criterion (eV/\AA)} & Energy cutoff (eV)\\ \hline 
bulk      & Mo$_{64}$Ta$_{64}$ &   1 & $4\times4\times4$ & $1\times 10^{-5}$ & $0.02$  & 224.6 \\
Ta$\to$Mo & Mo$_{65}$Ta$_{63}$ &  64 & $4\times4\times4$ & $1\times 10^{-5}$ & $0.02$  & 224.6 \\
Mo$\to$Ta & Mo$_{63}$Ta$_{65}$ &  64 & $4\times4\times4$ & $1\times 10^{-5}$ & $0.02$  & 224.6 \\
\hline \hline
\end{tabularx}
\end{threeparttable}
\end{table}

\clearpage
%%%%%%%%%%%%%%%%%%%%%%%%%%%%%%%%%%%%%%%%%%%
%%%%%%%%%%%%%%%%%%%%%%%%%%%%%%% Reference
\bibliographystyle{unsrt}
\bibliography{biblio.bib}